# Optical magnetic detection of single-neuron action potentials using quantum defects in diamond

J. F. Barry[1,2,3], M. J. Turner[2,3], J. M. Schloss[3,4], D. R. Glenn[1,2,3], Y. Song[5,6,7], M. D. Lukin[2], H. Park[2,3,8,9], R. L. Walsworth[1,2,3]*

[1] Harvard-Smithsonian Center for Astrophysics, Cambridge, Massachusetts 02138, USA.
[2] Department of Physics, Harvard University, Cambridge, Massachusetts 02138, USA.
[3] Center for Brain Science, Harvard University, Cambridge, Massachusetts 02138, USA.
[4] Department of Physics, Massachusetts Institute of Technology, Cambridge, Massachusetts 02139, USA.
[5] Marine Biological Laboratory, Woods Hole, Massachusetts 02543, USA.
[6] Yale School of Medicine, Department of Genetics and Howard Hughes Medical Institute, Boyer Center, New Haven, Connecticut 06510, USA.
[7] Boston Children's Hospital, Harvard Medical School, Department of Neurology/F.M. Kirby Neurobiology, Boston, Massachusetts 02115, USA.
[8] Department of Chemistry and Chemical Biology, Harvard University, Cambridge, Massachusetts 02138, USA.
[9] Broad Institute of MIT and Harvard, Cambridge, Massachusetts 02142, USA.
*Correspondence to: rwalsworth@cfa.harvard.edu

**A key challenge for neuroscience is noninvasive, label-free sensing of action potential (AP) dynamics in whole organisms with single-neuron resolution[1]. Here, we present a new approach to this problem: using nitrogen-vacancy (NV) quantum defects in diamond to measure the time-dependent magnetic fields produced by single-neuron APs. Our technique has a unique combination of features: (i) it is noninvasive, as the light that probes the NV sensors stays within the biocompatible diamond chip and does not enter the organism, enabling activity monitoring over extended periods; (ii) it is label-free and should be widely applicable to most organisms; (iii) it provides high spatial and temporal resolution, allowing precise measurement of the AP waveforms and conduction velocities of individual neurons; (iv) it directly determines AP propagation direction through the inherent sensitivity of NVs to the associated AP magnetic field vector; (v) it is applicable to neurons located within optically opaque tissue or whole organisms, through which magnetic fields pass largely unperturbed; and (vi) it is easy-to-use, scalable, and can be integrated with existing techniques such as wide-field and superresolution imaging. We demonstrate our method using excised single neurons from two invertebrate species, marine worm and squid; and then by single-neuron AP magnetic sensing exterior to whole, live, opaque marine worms for extended periods with no adverse effect. The results lay the groundwork for real-time, noninvasive 3D magnetic mapping[2] of functional neuronal networks, ultimately with synapse-scale (~10 nm) resolution[3] and circuit-scale (~1 cm) field-of-view[4].**

There are many established and emerging techniques for probing neuronal network activity, either at the 'micro-scale' with single-neuron resolution, or at the 'macro-scale' with whole organism compatibility. However, challenges remain to realize the complete set of desired capabilities[1] (see SI Table S1). In particular, electrophysiology recording methods such as patch-clamping remain the gold standard for measuring individual neuron action potentials (APs), with excellent signal-to-noise ratio (SNR) and good temporal resolution. Nonetheless, such direct-contact methods are technically delicate and invasive, and are not scalable to dense recording with both high spatial resolution and wide field-of-view[5]. Microelectrode arrays are more robust and less invasive, but spatial resolution is limited to ~10 μm and error-prone post-processing of data is required[6]. Optical techniques offer many advantages, but typically are limited to probing tissue depths ~1 mm due to scattering, and can employ optical power in excess of photo-damage thresholds[7]. In addition, calcium imaging is hampered by extraneous intracellular calcium[8], and its limited temporal resolution precludes resolving individual APs that fire at rates > 10 Hz[9]; voltage sensitive dyes have a tradeoff between poor SNR and high toxicity to the membrane[10]; functional near infrared spectroscopy (fNIRS) cannot resolve single-neuron activity; and voltage-sensitive fluorescent proteins must be genetically expressed[11], which may alter neuronal function[12].

Alternatively, AP _magnetic sensing_ confers important advantages: it is noninvasive, label-free, and able to detect neuronal activity through intervening tissue and whole organisms; and it provides inherent information on AP propagation direction and conduction velocity via the characteristic AP azimuthal vector magnetic field (see Fig. 1a). To date, however, magnetic techniques for sensing neuronal activity have either operated at the macro-scale with coarse spatial (~1 mm) and temporal (~1 ms) resolution — e.g., functional magnetic resonance imaging (fMRI) and magnetoencephalography (MEG) — or been restricted to biophysics studies of excised neurons probed with cryogenic or bulky detectors that are not scalable to functional networks or whole organisms[13].

As demonstrated here, the benefits of AP magnetic sensing can be realized with both single-neuron resolution and whole organism applicability using nitrogen-vacancy (NV) color centers in diamond. NV centers are atomic-scale quantum defects that provide nanoscale magnetic field sensing and imaging via optically detected magnetic resonance (ODMR), with broad applicability to both physical[14] and biological[4,15,16] systems under ambient conditions. We employ a simple, robust apparatus (see Fig. 1b) with a magnetic field sensor consisting of a macroscopic, single-crystal diamond chip with a uniform 13 μm layer containing a high density (~3×$10^{17}$ cm$^{-3}$) of NV centers at the top surface. The biological sample is placed on or above the NV-enriched surface. Laser light at 532 nm is applied to the sensing NV layer through the diamond at a sufficently shallow angle that the light reflects off the top diamond surface (due to total internal reflection) and therefore does not irradiate the living sample. Microwaves (MWs) are applied to the NV sensor via a wire loop located above the diamond, with minimal perturbation to the specimen studied[17]. Laser-induced fluorescence (LIF) from the NVs is imaged onto a photodiode; and continuous-wave electron spin resonance (CW-ESR) magnetometry is used to detect the AP magnetic field as a time-varying shift in the center of the ODMR spectrum, with temporal resolution of ~32 μs (see SI and SI Fig. S7). We regularly achieve magnetic field sensitivity $\eta = 15 \pm 1$ pT/$\sqrt{\text{Hz}}$ from a sensing volume of about $(13 \times 200 \times 2000)$ μm$^3$, which represents a twenty-fold improvement over previous broadband NV-diamond magnetometers[18] and provides SNR > 1 for a single AP event using matched filtering (see SI). For each biological specimen we typically acquire repeated AP magnetic field measurements, often over extended periods of time (hours). Multiple, consecutive AP measurements ($N_{\text{avg}}$) can also be averaged together to increase the AP SNR (see Methods and SI).

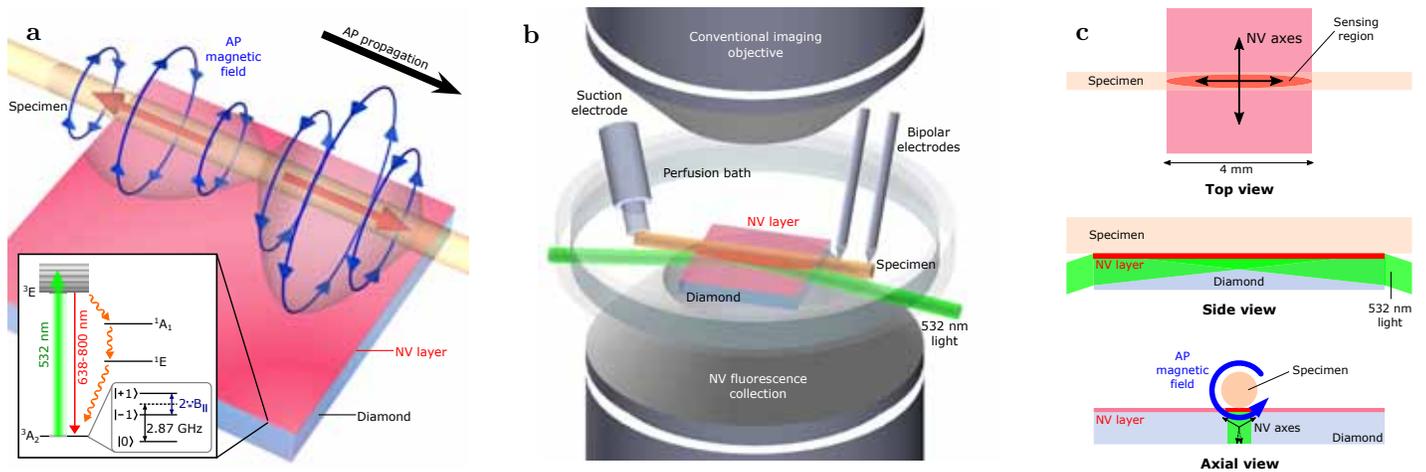

**Figure 1: Experimental overview. a,** Schematic image depicting bipolar azimuthal magnetic field associated with action potential (AP) propagating from left to right. Red arrows indicate axial current through axon and blue arrows depict associated magnetic field. Magnetic field projection is detected by 13 μm thick nitrogen-vacancy (NV) layer on diamond substrate. Inset shows NV center energy level diagram; see SI for details. **b,** Custom-built microscope allows simultaneous magnetic sensing and conventional imaging of specimens. NV centers are excited by 532 nm laser light oriented at grazing incidence to diamond top surface. Inverted aspheric condenser objective collects NV laser-induced fluorescence (LIF). Specimens are placed on top of diamond, and individual APs are stimulated by suction electrode and detected downstream via a pair of bipolar recording electrodes. For clarity, wire loop for microwave (MW) delivery and axon clamp are not shown. **c,** Top, side, and axial views of NV-diamond sensor and specimen. Top view shows sensing region from which LIF is collected, as well as four crystallograpic NV axes. AP magnetic field projects onto two NV axes perpendicular to specimen axis. Side view shows 532 nm laser light entering diamond at grazing angle and exciting NV layer. Blue arrow in axial view depicts AP magnetic field; black arrows depict NV axes in sensing region.

We first performed magnetic sensing of single-neuron APs from excised invertebrate giant axons, together with simultaneous electrophysiology measurements on the axons as a comparison and check on the magnetic data. We studied two species, with consistent results: the marine fanworm *Myxicola infundibulum*; and the North Atlantic longfin inshore squid *Loligo pealeii*, a model organism for neuroscience. Details of specimen preparation, AP stimulation, and electrophysiology measurements are described in the Methods. Figure 2a shows a representative measured intracellular AP voltage time trace $\Phi_{\text{in}}^{\text{meas}}(t)$ from *M. infundibulum*. In a simple model of the electromagnetic dynamics of APs (see SI), the magnetic field $B(t)$ is proportional to the temporal derivative of the intracellular voltage $\Phi_{\text{in}}(t)$: $B(t) = s d\Phi_{\text{in}}/dt$, where $s$ is a scaling constant dependent on geometrical parameters (axon radius $r_a$, radial distance of the field point to the axon center $\rho$) and electrophysiological axon parameters (AP conduction velocity $v_c$, axoplasm electrical conductivity $\sigma$). As shown in Figs. 2b and 2c, we find good agreement between (i) $B^{\text{calc}}(t)$, the AP magnetic field calculated from $\Phi_{\text{in}}^{\text{meas}}(t)$ for a typical value of $s$ for *M. infundibulum*, and (ii) a representative measured AP magnetic field time trace $B^{\text{meas}}(t)$. This correspondence demonstrates the consistency of NV-diamond magnetic AP measurements with standard electrophysiology techniques and theory. Note that this example $B^{\text{meas}}(t)$ data has a peak-to-peak amplitude $= 4.1 \pm 0.2$ nT for $N_{\text{avg}} = 150$, corresponding to an SNR of $1.2 \pm 0.1$ for a single AP firing, i.e., $N_{\text{avg}} = 1$ (see SI). Using $N_{\text{avg}} = 6$ yields an SNR of 3, which is sufficient for AP event detection. Furthermore, we demonstrated that our method has multi-species capability via magnetic sensing of APs from the squid *L. pealeii* (Fig. 2d). No change to the apparatus or magnetic sensing protocol is required upon switching organisms.

We next demonstrated single-neuron AP magnetic sensing exterior to a whole, live, opaque organism — an undissected specimen of *M. infundibulum* (Fig. 3a) — for extended periods, with minimal adverse effect on the animal (see SI and SI Fig. S8b). For example, Fig. 3b shows an example measured AP magnetic field time trace $B^{\text{meas}}(t)$ for a live intact specimen, which is to the best of our knowledge the first demonstration of 'single- neuron MEG' from the exterior of a whole animal. The measured AP waveform in Fig. 3b is similar to that of an excised axon (Fig. 2c), with roughly four times smaller peak-to-peak amplitude ($\approx 1$ nT), which is consistent with the separation of ~1.2 mm from the center of the axon inside the animal to the NV sensing layer (see transverse sections and diagrams in Fig. 4a-d and in SI Fig. S1d, and SI). In addition, we recorded $B^{\text{meas}}(t)$ from a live intact worm after > 24 hours of continuous exposure to the experimental conditions, including applied MWs and optical illumination of the diamond sensor. We observed little to no change in the magnetic AP signal or in the animal behavior (see SI Fig. S8b and SI).

We also used whole live worms to demonstrate the capability of NV-diamond magnetic sensing to determine the AP propagation direction and distinguish differences in AP conduction velocity ($v_c$) from a single-point measurement. NV-diamond provides full vector magnetometry by sensing the magnetic field projection onto a linear combination of the four NV center orientations within the diamond crystal lattice (see SI). A neuron AP produces a bipolar azimuthal magnetic field waveform, with the time-varying field orientation set by the direction of AP propagation (see Figs. 1a and 4d). Thus, as shown in Fig. 4e, f, the measured AP magnetic field time trace $B^{\text{meas}}(t)$ from an intact worm has an inverted waveform for anterior versus posterior AP stimulation, demonstrating clear distinguishability between oppositely-propagating APs. Furthermore, the amplitude of the AP magnetic field, at a given radial distance from an axon, is expected from cable theory (see SI) to scale inversely with AP conduction velocity (i.e., as $\sim 1/v_c$), which could enable NV-diamond magnetometry as a sensitive probe of axon demylination as well as other neurophysiological effects affecting $v_c$. For example, the giant axon radius in *M. infundibulum* is tapered over the organism's length[19], which is predicted to induce a propagation-direction-dependent asymmetry in $v_c$[20]. Specifically, $v_c$ is expected to be smaller, and thus $B(t)$ to be larger, for an AP propagating in the direction of positive taper (increasing axon radius) than for an AP propagating in the direction of negative taper (decreasing radius). Our measurements in whole worms are consistent with this prediction: transverse sections show a taper in the axon radius from smaller near the posterior to larger near the anterior (Fig. 4a, b), correlated with a larger amplitude of $B^{\text{meas}}(t)$ at a fixed measurement point along the axon for posterior versus anterior AP stimulation (Fig. 4e, f). We observe this asymmetry in all three worms tested in this way, with amplitude differences of 47% ± 20% (see SI Fig. S8a). Independent two-point electrophysiology measurements (see SI) provide consistent results, with smaller measured $v_c$ (ratio ≈ 0.6) for posterior versus anterior stimulation.

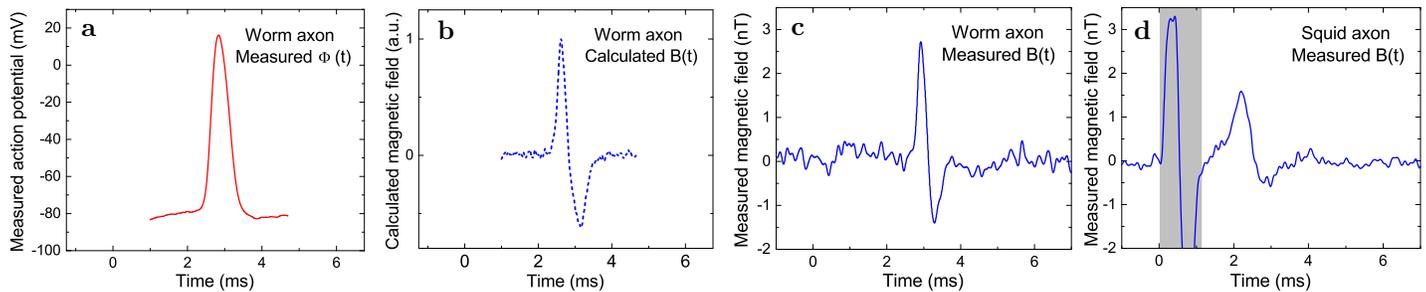

**Figure 2: Measured action potential voltage and magnetic field from excised single-neurons. a,** Measured time trace of action potential (AP) voltage $\Phi_{\text{in}}^{\text{meas}}(t)$ for giant axon from *Myxicola infundibulum* (worm). **b,** Calculated time trace of AP magnetic field $B^{\text{calc}}(t)$ for *M. infundibulum* extracted from data in Fig. 2a. **c,** Measured time trace of AP magnetic field $B^{\text{meas}}(t)$ for *M. infundibulum* giant axon with $N_{\text{avg}} = 150$. **d,** Measured time trace of AP magnetic field $B^{\text{meas}}(t)$ for *Loligo pealeii* (squid) giant axon with $N_{\text{avg}} = 375$. Gray box indicates magnetic artifact from stimulation current.

It is worth highlighting that, in contrast to our method of <u>vector</u> magnetic field sensing using NV-diamond, existing techniques for AP detection with single-neuron resolution measure a <u>scalar</u> quantity, e.g. ion concentration or electric potential. Such scalar measurements do not provide single-point determination of AP conduction velocity magnitude or direction; instead, spatially separated differential scalar measurements must be performed, with sufficiently high temporal resolution and SNR to distinguish small timing differences between the two detected signals. Specifically, when the AP spatial extent is long compared to the axon (as in many systems of interest in neuroscience), the delay between the detected scalar signals at two measurement points along the axon is significantly shorter than the duration of each signal. As a result, determining AP conduction velocity via scalar techniques requires a much higher SNR than with vector magnetometry.

Building on our present results, NV-diamond magnetic sensing should be applicable to noninvasive monitoring of AP activity in a broad range of systems such as cultured neurons, tissue slices, and whole organisms, including species for which genetic encoding and viral transduction of voltage-sensitive proteins is not currently possible. An example near-term application is single-point measurements of AP conduction velocity, which could greatly aid the study of demyelinating diseases such as multiple sclerosis. Furthermore, NV-diamond magnetic sensing could be combined with optical stimulation methods[21] to provide individual-neuron-targeted excitation and noninvasive AP detection, enabling longitudinal studies of environmental and developmental effects, and tests of models used to interpret conventional MEG signals from macroscopic brain circuits. Key technical challenges for neuroscience applications of NV-diamond include: (i) improving the magnetic field sensitivity to enable real-time, single AP event detection from individual mammalian neurons, which are expected to generate peak AP magnetic fields ~1 nT at the NV sensor layer (see SI); and (ii) incorporating magnetic imaging. The imaging challenge can be met by integrating techniques from our recent successful demonstrations using NV-diamond for wide-field parallel magnetic imaging of biological cells[4,15], and superresolution magnetic imaging[3], as well as tomographic methods for extending the depth-of-field[2]. The sensitivity challenge can be addressed by using optimized diamonds with higher NV density and longer spin-dephasing times $T_2^*$, and by implementing pulsed-Ramsey[22] and quantum-beat[23] measurement protocols; a sensitivity and AP SNR gain of ~$10^4$ per unit sensor volume is expected, along with temporal resolution down to ~1 μs (see SI). To realize further sensitivity enhancements we will investigate quantum-assisted techniques, which should enable measurements approaching fundamental quantum limits. Our present NV-diamond instrument has a photon-shot-noise-limited magnetic field sensitivity ~3000 times worse than the quantum spin-projection limit (see SI), highlighting the potential for large sensitivity gains. For example, we recently demonstrated that spin-to-charge-state readout for NV centers provides enhanced magnetic field sensitivity that is only a factor of 3 above the spin-projection limit[24].

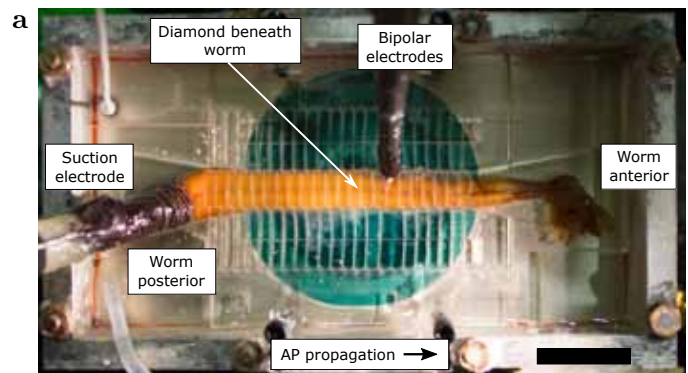

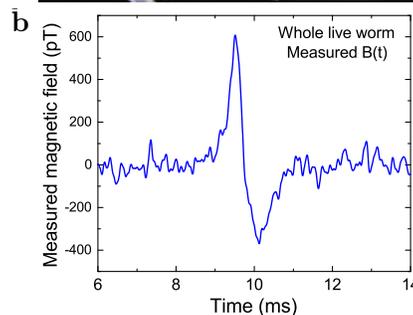

**Figure 3: Single-neuron action potential magnetic sensing exterior to whole live organism. a,** Overhead view of intact living specimen of *M. infundibulum* (worm) on top of NV-diamond sensor. In configuration shown, animal is stimulated from posterior end by suction electrode; action potentials (APs) propagate toward worm's anterior end; and bipolar electrodes confirm AP stimulation and propagation. Scale bar is 20 mm. **b,** Recorded time trace of single-neuron AP magnetic field $B^{\text{meas}}(t)$ from live intact specimen of *M. infundibulum* for $N_{\text{avg}} = 1650$ events.

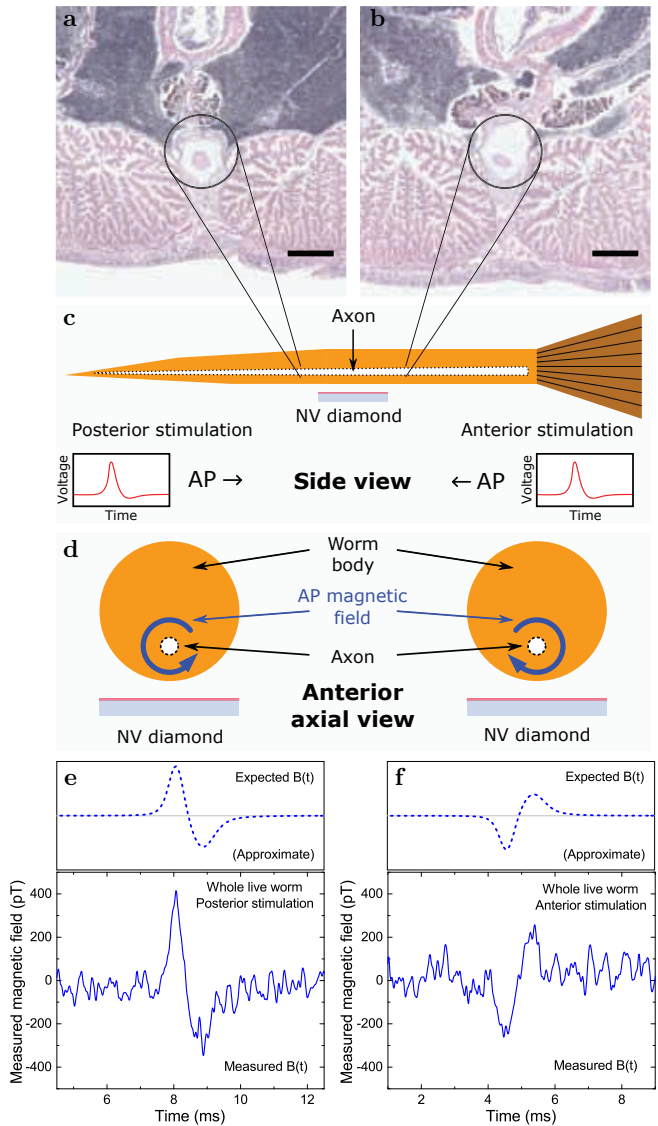

**Figure 4: Single-point sensing of action potential propagation direction and conduction velocity exterior to whole live organism.** Transverse sections of *M. infundibulum* near mid-point of worm illustrate giant axon radius tapering from smaller near posterior (**a**) to larger near anterior (**b**). Sections taken ~1 cm apart. Encircled white structure is giant axon. Scale bars are 400 μm. **c,** Cartoon cross-section side view of whole live worm and NV-diamond sensor. Black dashed lines indicate tapered giant axon. Cartoon time traces of action potential (AP) voltage indicate they are typically identical for posterior stimulation (left-propagating AP) and anterior stimulation (right-propagating AP). **d,** Cartoon cross-section axial view looking from anterior end. Blue arrows encircling axon indicate opposite azimuthal AP magnetic field vectors for oppositely propagating APs. **e,** Top: Expected AP magnetic field time trace for posterior worm stimulation, indicating effect of AP propagation direction and conduction velocity on sign of bipolar magnetic field waveform and magnetic field amplitude, respectively. Bottom: Recorded time trace of AP magnetic field $B^{\mathrm{meas}}(t)$ from intact live specimen of *M. infundibulum* for posterior stimulation and $N_{\mathrm{avg}} = 1650$ events. **f,** Top: Expected AP magnetic field time trace for anterior worm stimulation. Bottom: Recorded time trace of AP magnetic field $B^{\mathrm{meas}}(t)$ from the same intact live specimen of *M. infundibulum* as in (e) for anterior stimulation and $N_{\mathrm{avg}} = 1650$ events. Note that the observed sign of $B^{\mathrm{meas}}(t)$ is reversed depending on AP propagation direction; and the magnetic signal amplitude is larger by about 50% for posterior-stimulated APs, consistent with two-point electrophysiology measurements of smaller AP conduction velocity for posterior stimulation.

Ultimately, we envision that NV centers or other quantum defects in diamond[25] will enable a qualitatively new imaging modality for neuroscience — a 'quantum diamond microscope' that provides fast (~1 μs), real-time, 3D magnetic mapping of functional activity, connectivity, and directionality in neuronal networks, including in whole organisms and with single-neuron resolution; and which is integrable with superresolution methods to yield ~10 nm spatial resolution[3], or with wide-field imaging[4] and tomographic techniques[2] to study tissue volumes ~1 cm³. This new functional imaging modality could be applied to detailed studies of the relationship between microscopic neuronal connectivity and circuit development and function: e.g., the gain and loss of branches and synapses as well as the adaptive strengthening and weakening of connections at the level of single cells, as governed by intrinsic neuronal cell behaviors and spatiotemporal patterns of neuronal signals and biochemical rewards, and the resulting effect on and feedback from overall circuit function[26,27]. At the extremes of technical performance, one might also observe opening and closing events of single ion channels[28], detect spatial heterogeneity in the radial AP currents due to nodes of Ranvier[29], and map sub-threshold currents in neural dendrites and soma[30].

**Acknowledgements |** We thank Element Six for diamond samples used in this work; E. R. Soucy and J. Greenwood for machining assistance and advice on electrophysiology techniques; S. G. Turney, M. G. Shapiro, P. Ramesh, and H. Davis for technical guidance and fruitful discussions; D. Bowman and B. Faulkner-Jones of the Dept. of Pathology at Beth Israel Deaconess Medical Center for tissue processing and imaging of transverse sections; and the Center for Brain Science at Harvard University for infrastructure and support. J.M.S. was supported by a Fannie and John Hertz Foundation Graduate Fellowship and a National Science Foundation Graduate Research Fellowship under Grant No. 1122374. This work was financially supported by the DARPA QuASAR program, the MURI biological transduction program, the NSF, and the Smithsonian Institution.

## Methods

### *M. infundibulum* properties and acquisition

*M. infundibulum* specimens possess a large giant axon[19], are readily available year-round, and can be kept for long periods of time in laboratory environments. The giant axon mediates a rapid escape reflex; electrical or physical stimulus elicits a violent muscular contraction, which can shorten the worm by 50% or more compared to the relaxed state. Specimens are obtained from a commercial supplier (Gulf of Maine Inc., Bay of Fundy, Maine, USA) or a research laboratory (UC Davis Bodega Bay Marine Biology Laboratory, California, USA). Worms are housed in a 208 L aquarium filled with artificial seawater (ASW) from a commercial supplier (Instant Ocean Sea Salt), and temperature stabilized to 7.5 ± 0.5 °C. Worms are fed a plankton-based food source (Sera Marin Coraliquid) every 14 days. Large specimens of length ~60 mm and diameter > 5 mm (both measured when fully contracted) were used in the present studies.

### *M. infundibulum* specimen preparation and action potential stimulation

For studies of the excised giant axon of *M. infundibulum*, a modified version of the Binstock and Goldman method was followed[31]. (i) The ventral side of *M. infundibulum* is identified by a triangular structure on the thorax[32]. (ii) The worm is pinned ventral side down in a glass dish spray-painted flat black (Kyrlon Ultra Flat Black #1602) and filled with PDMS (Dow Corning Sylgard 184). The specimen is illuminated with broadband white light at a shallow grazing angle. The painted dish and lighting increase contrast between the axon and the surrounding tissue for better visibility. The preparation is submerged in chilled ASW throughout. (iii) A median dorsal incision of length ~2 cm is made through the body wall at the mid-section of the animal. Further cuts sever connective tissue between the body wall and the gut. (iv) The freed body wall is pinned to the PDMS away from the axon with substantial tension as described in ref. 31. (v) The gut is partially lifted (vertically up) from the body wall, allowing fine vanassas scissors to cut the connective tissue connecting the gut to the body wall. The gut is excised from the specimen, revealing the dorsal side of the ventral nerve cord containing the giant axon. (vi) The ventral blood vessels and additional tissue close to the axon are carefully stripped away with fine forceps, further exposing the nerve cord as shown in SI Fig. S1a, b. (vii) Additional cuts through the body wall remove tissue around the nerve cord as shown in SI Fig. S1c. (viii) The excised nerve cord (still connected to the undissected worm anterior and posterior) is then placed dorsal-side-down on the diamond sensor chip. In this configuration the worm is alive, and action potential (AP) firing can continue for 72 hours or longer. A flexible acrylic clamp holds the nerve cord fixed against the diamond chip to restrain the worm's muscle contractions. Axon viability is checked periodically through physical or electrical stimulus at the worm posterior and confirmation of muscle response at the worm anterior. Note that for studies of <u>intact</u> specimens of *M. infundibulum*, the worm is cleaned of accumulated mucus and placed in an acrylic jig to fix the dorsal side of the animal against the diamond. For both excised axons and whole live worms, the specimen preparation is continuously perfused with 10 °C ASW with a gaseous solute of 99.5% $O_2$ and 0.5% $CO_2$.

APs are evoked for duration $\Delta t_{stim}$ by a suction electrode engaging either the specimen posterior or anterior, and driven by an isolated pulse stimulator (A-M Systems Model 2100) producing biphasic pulses (positive polarity first) with 10 mA amplitude. Posterior stimulation is used unless otherwise noted. The value of $\Delta t_{stim}$ is typically set to twice the stimulation threshold, and ranges from 100 µs to 1 ms depending on the size and health of the organism and the degree of contact between the worm body and the suction electrode. Stimulation pulses are applied at repetition rate $f_{stim} \approx 0.4$ Hz. Successful AP stimulation and propagation are verified with a pair of bipolar platinum recording microelectrodes (World Precision Instruments PTM23B10 or PTM23B05) connected to a differential amplifier (A-M Systems Model 1800 Headstage), which is further amplified (A-M systems Model 1800) and then digitized (National Instruments USB-6259) at 250 kHz. For the intact worm studies, the same recording electrodes were gently positioned in contact with the worm skin near the axon, allowing verification of AP propagation. Care was taken to not puncture the skin to avoid damaging the specimen.

### *L. pealeii* acquisition, preparation, and action potential stimulation

Specimens of *L. pealeii* are acquired seasonally from the Marine Biological Laboratory in Woods Hole, Massachusetts, USA, with medium to large squid (0.3 m to 0.5 m in overall length) chosen for the present studies. Following decapitation of the squid, the post-synaptic giant axons are isolated following the protocol described in ref. 33. The axons are placed in vials containing calcium-free saline solution and stored on ice. The iced vials are transported from Woods Hole to our laboratory at Harvard University (90-minute drive). The isolated axons maintain viability for up to 12 hours after initial excision.

For studies of the excised giant axon of *L. pealeii*, a squid buffer solution (475 mM NaCl, 115 mM $MgCl_2$, 10 mM $CaCl_2$, 10 mM KCl, 3 mM $NaHCO_3$, 10 mM HEPES) chilled to 10 °C perfuses the axon. AP stimulation and extracellular voltage detection is accomplished through the same methods used for *M. infundibulum*, save for an increase in the stimulation repetition rate $f_{stim}$. The squid giant axon can be fired as often as 100 Hz without reducing detected AP signal quality. For increased longevity of squid axons, stimulation is applied at $f_{stim} = 25$ Hz for 15 s followed by a rest period of 15 s.

### Electrophysiology

Determination of $B^{calc}(t)$ requires recording the intracellular AP voltage $\Phi_{in}(t)$. Intracellular microelectrodes are pulled from commercial glass (World Precision Instruments 1B150-4) to a resistance of 20 - 40 MΩ on a pipette puller (Dagan DMZ Universal Puller), filled with 3M KCl, and fitted into a headstage (Axon Instruments HS-2A) connected to an amplifier (Axon Instruments AxoProbe 1B). The microelectrode is mounted to a micro-manipulator for precise insertion into the axon. *M. infundibulum* specimens are dissected as described previously but remain pinned in the dissection dish during this measurement so that the microelectrode remains sealed to the axon despite the worm's contractions. AP stimulation occurs as described previously. The time trace $\Phi_{in}^{meas}(t)$ is digitized (Tektronix TDS2004B), and subsequently low-pass-filtered at 2 kHz. Axon resting potential values $\Phi_0$ of -60 mV to -85 mV and peak AP amplitudes of 100 mV to 110 mV are observed for *M. infundibulum*, consistent with typical values in the literature[31].

For *M. infundibulum*, APs were found to be abolished for values of $f_{stim} \gtrsim 5$ Hz, and maximal AP amplitudes realized for $f_{stim} \lesssim 1$ Hz, both in agreement with previous reports in the literature[34]. For *L. pealeii*, consistent AP stimulation was observed up to $f_{stim} = 300$ Hz, although eventual axon degradation was observed over ~15 minutes at such high rates. *L. pealeii* exhibits a refractory period following the AP, wherein the potential experiences an overshoot below the resting voltage. The overshoot is not present in *M. infundibulum*.

**Transverse sections**
Transverse sections shown in Fig. 4a, b were prepared from a typical *M. infundibulum* specimen, which was anesthetized and fixed in a solution of 4% paraformaldehyde (PFA) overnight. The fixed worm was mounted to a paraffin block and sectioned, yielding slices of width ∼4 μm. The slices were treated with a hematoxylin and eosin stain (H&E stain) to illustrate the tissue structure. Representative slices from three different sections are shown (https://slide-atlas.org/link/mn74xw). By comparison with the literature[19], we verified the structure observed in the sections and confirmed the location of the giant axon. The sections show a distance from the giant axon center to the skin surface of $900 \pm 200$ μm, and a taper in the axon with decreasing diameter from anterior to posterior.

## Supporting Information

### NV-diamond physics and magnetometry method

NV colour centers are localized quantum defects in diamond consisting of a substitutional nitrogen adjacent to a vacancy in the lattice. The NV center has an $S = 1$ (triplet) ground state with a zero-field splitting of 2.87 GHz between the $m_s = 0$ and $m_s = \pm 1$ spin projections. These states have additional hyperfine structure, which arises from the coupling of the $^{14}$N nuclear spin $I = 1$ to the unpaired NV electron spin. A local magnetic field induces Zeeman shifts, lifting the degeneracy of the $m_s = \pm 1$ energy levels. Optically-induced electronic transitions to the excited triplet state and fluorescent decay back to the ground electronic state are mainly spin-conserving[35]. Fluorescent readout and optical polarization of the NV spin state are made possible through a nonradiative decay path from the $m_s = \pm 1$ excited states through metastable singlet states and preferentially to the $m_s = 0$ state[35].

The NV spin is quantized along one of four crystallographic NV symmetry axes in the diamond crystal, which are equally populated for a typical ensemble of NV centers[36]. Thus the diamond sensor chip used in the present work, containing a large ensemble of NV centers, provides a complete basis for vector magnetometry: a local magnetic field's magnitude and direction can be reconstructed from its measured projections onto each of the NV axes[37,15]. AP magnetic fields are expected to encircle the axon and be directed perpendicular to the axon axis and hence the direction of AP propagation. In this work the axons were oriented roughly linearly on the diamond and normal to two NV axes, maximizing the projection of the AP magnetic field $B(t)$ onto those axes, as shown in Fig. S2a, with the projection along the other two NV axes expected to be near-zero. In this configuration the component of $B(t)$ parallel to the NV surface layer and perpendicular to the axon axis is detected. In future incarnations of an instrument for magnetic imaging of networks of neurons oriented arbitrarily on the diamond surface, the component of $B(t)$ perpendicular to the NV layer at each point on the diamond surface would be sensed, as shown in Fig. S2b. The magnetic field projection onto a single NV axis would have opposite sign for measurement points on different sides of the axon, and $B(t)$ would in general have nonzero projection on each of the four NV axes.

A modified CW-ESR technique is employed for NV-diamond ODMR, wherein optical NV spin polarization, MW drive, and spin-state readout via LIF occur simultaneously. Continuous green laser excitation at 532 nm polarizes the NV center into the $m_s = 0$ ground state. Applied MWs, when tuned to resonance with the transition between the optically bright $m_s = 0$ spin state and one of the less bright $m_s = +1$ or $-1$ states, cause NV spin precession into a mixed state and a detectable reduction in LIF. A change in the local magnetic field shifts the ODMR feature and, for near-resonant MW drive, is detected as a change in the fluorescence rate.

A single ODMR feature of Lorentzian lineshape with angular frequency $\omega_0$ (where $\omega \equiv 2\pi f$), linewidth $\Gamma$, and contrast $\mathcal{C}$ is detected in fluorescence as $F(\omega) = F_0\left(1 - \mathcal{C}\frac{(\Gamma/2)^2}{(\Gamma/2)^2 + (\omega - \omega_0)^2}\right)$ (see Fig. S3a), ignoring MW and optical broadening here for simplicity. As the majority of noise in the system has $1/f$ character, greater SNR is achieved by shifting the measurement bandwidth to higher frequency via a lock-in technique, which generates a dispersion-like signal with a characteristic zero-crossing feature: i.e., a rapid change of the lock-in amplifier (LIA) voltage and sign with frequency. The applied MWs are square-wave frequency modulated at $f_{\text{mod}}$ (typically 18 kHz) about the center frequency $\omega_c$ with frequency deviation $\omega_{\text{dev}}$: i.e., $\omega_{\text{MW}}(t) = \omega_c + \omega_{\text{dev}}\text{square}(2\pi f_{\text{mod}} t)$. The collected fluorescence is then $F(\omega_{\text{MW}})$. After demodulation by the LIA with a reference signal $V_{\text{ref}}\sin(2\pi f_{\text{mod}} t)$, the DC output is a dispersion-type signal with a zero-crossing at $\omega_0$:

$$V_{\text{LIA}}(\omega_c, \omega_{\text{dev}}) \propto \frac{F(\omega_c + \omega_{\text{dev}}) - F(\omega_c - \omega_{\text{dev}})}{2}$$
$$= \frac{V_0 \mathcal{C}}{2}\left(-\frac{(\Gamma/2)^2}{(\Gamma/2)^2 + ((\omega_c + \omega_{\text{dev}}) - \omega_0)^2}\right.$$
$$\left.+ \frac{(\Gamma/2)^2}{(\Gamma/2)^2 + ((\omega_c - \omega_{\text{dev}}) - \omega_0)^2}\right),$$

where $V_0$ is a prefactor voltage determined by $F_0$ and the output settings of the LIA. Setting $\omega_{\text{dev}} = \frac{\Gamma}{2\sqrt{3}}$ maximizes the slope of the zero-crossing $\left.\frac{dV_{\text{LIA}}}{d\omega_c}\right|_{V_{\text{LIA}}=0}$ in the absence of power broadening[38]. Time-varying magnetic fields $B(t)$ are sensed by setting $\omega_c = \omega_0|_{t=0}$ and detecting resonance frequency shifts $\omega_0(t) = \omega_0 + \delta\omega(t)$, where $\delta\omega(t) = \frac{g_e \mu_B}{\hbar}B(t)$, as:

$$V_{\text{LIA}}(t) = V_{\text{LIA}}\left(\omega_0 - \frac{g_e \mu_B}{\hbar}B(t)\right)$$
$$= \frac{V_0 \mathcal{C}}{2}\left(-\frac{(\Gamma/2)^2}{(\Gamma/2)^2 + \left(\frac{\Gamma}{2\sqrt{3}} - \frac{g_e \mu_B}{\hbar}B(t)\right)^2}\right.$$
$$\left.+ \frac{(\Gamma/2)^2}{(\Gamma/2)^2 + \left(\frac{\Gamma}{2\sqrt{3}} + \frac{g_e \mu_B}{\hbar}B(t)\right)^2}\right)$$
$$\approx -\frac{3\sqrt{3}}{4}\frac{V_0 \mathcal{C}}{\Gamma}\frac{g_e \mu_B}{\hbar}B(t).$$

The NV spin resonance has three features separated by the hyperfine (HF) splitting of $\Delta\omega_{HF} = 2\pi \times 2.16$ MHz, as shown in Fig. S3b. For a single MW frequency sweeping across the features, and again ignoring MW power broadening, we find

$$F(\omega) = F_0\left(1 - \sum_{q=-1}^{1} \mathcal{C}\frac{(\Gamma/2)^2}{(\Gamma/2)^2 + (\omega - (\omega_0 + q\Delta\omega_{HF}))^2}\right).$$

Addressing all three NV HF features simultaneously with three MW frequencies also separated by $\Delta\omega_{HF}$ yields
$F(\omega, \Delta\omega_{HF})$
$$= F_0\left(1 - \sum_{p=-1}^{1}\sum_{q=-1}^{1}\mathcal{C}\frac{(\Gamma/2)^2}{(\Gamma/2)^2 + ((\omega + p\Delta\omega_{HF}) - (\omega_0 + q\Delta\omega_{HF}))^2}\right).$$

As displayed in Fig. S3c, the observed NV fluorescence signal shows five ODMR peaks. The outer two peaks correspond to one of the three MW frequencies on resonance; the second and fourth peaks correspond to two of the three frequencies tuned to resonance; and the innermost peak corresponds to all three MW frequencies resonantly addressing the HF features. The dispersion signal is then:
$V_{\text{LIA}}(\omega, \Delta\omega_{HF}, \omega_{\text{dev}})$
$$= V_0\left(\sum_{p=-1}^{1}\sum_{q=-1}^{1} -\mathcal{C}\frac{(\Gamma/2)^2}{(\Gamma/2)^2 + ((\omega + p\Delta\omega_{HF} - \omega_{\text{dev}}) - (\omega_0 + q\Delta\omega_{HF}))^2}\right.$$
$$\left.+ \mathcal{C}\frac{(\Gamma/2)^2}{(\Gamma/2)^2 + ((\omega + p\Delta\omega_{HF} + \omega_{\text{dev}}) - (\omega_0 + q\Delta\omega_{HF}))^2}\right).$$

In this simple treatment in which MW power broadening is ignored, our measurement technique increases the contrast of the central NV HF feature by a factor of 3. In practice, a contrast improvement factor of $\approx 1.9$ is achieved compared to the case of addressing a single HF feature.

The overall measurement contrast is further improved by orienting the bias field $B_0$ to have equal projections along two NV axes. Projecting along two NV axes doubles the contrast as shown by comparing Fig. S3c and d, although the angle between the NV axes and $B(t)$ causes the sensitivity improvement to be $2\cos[\pi/2 - \theta_{\text{tet}}/2]$ where $\theta_{\text{tet}} = 109.4712°$ is the tetrahedral bond angle in the diamond lattice.

**NV-diamond magnetometer details**

The diamond used in this work is an electronic grade (N < 5 ppb) single crystal chip, with rectangular dimensions 4 mm x 4 mm x 500 μm, grown using chemical vapor deposition (CVD) by Element Six. The 13 μm thick top-surface NV sensing layer consists of 99.99% $^{12}$C with 25 ppm $^{14}$N, which was irradiated with 4.6 MeV electrons with $1.3 \times 10^{14}$ cm$^{-2}$s$^{-1}$ flux for 5 hours and subsequently annealed in vacuum at 800 °C for 12 hours. The measured nitrogen-to-NV conversion efficiency is ~7%. The diamond is cut so that the 500 μm x 4 mm faces are perpendicular to the [110] crystal axis. The sides are mechanically ground to an optical-quality polish. The diamond is mounted to a 2" diameter, 330 μm thick silicon carbide (SiC) heat spreader via thermal epoxy (Epotek H20E) as shown in Fig. S5a. A 2 mm x 25 mm slot in the SiC provides access to the diamond surface for the dissected axon studies.

For NV magnetometry, the diamond sensor is illuminated by 2.75 - 4.5 W of 532 nm laser light (Coherent Verdi V-5) as shown in Fig. S4b. Laser light is guided into the diamond via an in-house fabricated UV fused silica coupler, making a ~ 3° angle to the NV layer. A 13 μm thick and 1 mm wide reflective aluminum layer on the diamond surface blocks both excitation light scattered by surface defects and LIF from impinging upon the specimen. A rare earth magnet (1" x 1" x 1" N42 K&J Magnetics) with south pole facing the experiment creates a bias magnetic field $B_0$ with equal projections of 7 gauss along the two NV axes normal to the axon axis, shifting the MW resonance between the $m_s = 0$ and $m_s = 1$ sublevels to ≈ 2.89 GHz. Figure S4a shows a schematic of the MW setup. A commercial MW source (Agilent E8257D) outputs a single near-resonant frequency, which is square-wave modulated at frequency $f_{\text{mod}} = 18$ kHz with frequency deviation $\omega_{\text{dev}} = 2\pi \times 360$ kHz (Rigol DG1022U). The modulated MWs pass through an isolator (Teledyne Microwave T-2S73T-II) and a -10 dB coupler before mixing (RELCOM M1G) with a 2.16 MHz sinusoid waveform (Stanford Research Systems DS345). The coupled port of the -10 dB coupler is further attenuated by 6 dB and combined (Mini-Circuits ZX-10-2-42-S+) with the mixer output and then sent through a second -10 dB coupler. The coupled output is sent to a spectrum analyzer (Agilent E4405B) while the transmitted output is amplified (Mini-Circuits ZHL-16W-43+), passed through another isolator (Teledyne Microwave T-2S73T-II), a circulator (Pasternack, PE 8401), and a high-pass filter (Mini-Circuits VHF-1200), before delivery to a square 5 mm x 5 mm loop located ≈ 2 mm above the diamond sensor. Slow variations in the NV ODMR resonances, e.g., due to diamond temperature drift, are compensated with ≈ 0.4 Hz feedback to the MW frequency $f_{\text{MW}}$. The ODMR features can also be used to continuously monitor the diamond substrate temperature in real time.

Typically 17 mW (and up to 28 mW) of LIF from the NV-diamond is collected by a 1.4 numerical aperture (NA) aspheric aplanatic oil condenser (Olympus), passed through a 633 nm long-pass filter (Semrock LP02-633RU-25), and imaged onto a biased photodiode (Thorlabs DET100A). The photodiode (PD) is powered by a 12 V lithium ion rechargable battery and is terminated into the RF + DC port of a bias tee (Universal Microwave Component Corporation BT-1000-LS) with bandwidth 10 kHz - 1 GHz. The bias tee DC port is terminated by 50 Ω during data taking; during optical alignment the port is monitored on an oscilloscope to optimize LIF collection. The RF output of the bias tee is amplified by a low noise amplifier (RF Bay LNA-545) and then sent into an LIA (Stanford Research System SR850). The LIA gain setting is 200 mV, and the nominal time constant is 30 μs with a 24 dB/octave roll-off, yielding a measured 3 dB cutoff frequency of $f_c = 3.6$ kHz and a measured equivalent noise bandwidth (ENBW) of $f_{\text{ENBW}} = 4.0$ kHz. The LIA voltage output is expanded by 5 times using the LIA expand function, digitized (National Instruments USB-6259) at 250 kHz, and then subsequently divided by 5. The temporary LIA signal expansion was found to reduce the effect of read noise from digitization. A 3 nT magnetic field corresponds to a fractional change in the NV LIF of $\Delta F/F \approx 1.4 \times 10^{-6}$.

To suppress laser intensity noise near $f_{\text{mod}}$, the 532 nm laser light is sampled and focused on a separate, reference PD (see Fig. S4b, c). This PD and all electronics (bias-tee, low-noise amplifier, LIA, input into data acquisition system) exactly duplicate the setup of the signal PD and accompanying electronics. The phase of the reference LIA is aligned with the signal LIA phase. We find subtraction (rather than division) of the correlated noise is sufficient to reach the photon shot noise sensitivity sensitivity limit in the absence of the MWs, in agreement with ref. 39. The detected signal is digitally filtered with a 80 Hz FFT high-pass filter; and 1-Hz-wide notch stop filters at all 60 Hz harmonics through 660 Hz and at 30 other frequencies above 2 kHz. The experiment achieves sensitivity ~50% above the photon-shot-noise limit, which is discussed in the sensitivity section below.

For the intact worm studies, several changes were made to the experimental apparatus (see Fig. S5b). An upgraded aluminum mount (larger than the mount for excised axons) is used to fit the large intact specimens (see Fig. 3a). A SiC wafer with no slot is used as a heat spreader. The NV-diamond sensor is therefore offset from the worm exterior by a spacer of thickness 330 μm. MWs are delivered to a 25 μm thick copper foil layer directly on top of the diamond. The Olympus oil aspheric condensor is exchanged for a 0.79 NA air aspheric condensor (Thorlabs ACL25416U-B). Stained transverse sections in Fig. 4a, b show a typical tissue thickness of ~900 μm from the center of the axon to the worm exterior, consistent with the literature[19,40], although this distance is also noted[19] to be highly variable among different specimens and along a single organism's length. The overall typical distance from the axon center to the diamond sensor is ~1.2 mm, consistent with the measured roughly four-fold magnetic signal reduction compared to excised worm axons, where the distance from axon center to NV detector layer is typically ~300 μm.

Excitation-laser-induced heating of the diamond is measured via NV ODMR frequency shifts to be 2.4 °C/Watt. For the data shown in Fig. 2c, d (Worm A and squid), the diamond temperature is $21 \pm 3$ °C. As the excised axons are placed directly against the diamond, we estimate the temperature of both the Worm A and squid axons to be ~ 21 °C. The live intact organisms (Worm B of Fig. 3b and Worms C, D, E, and F of Fig. S8) are separated from the diamond by the SiC heat spreader and are thus at ~ 10 °C during sensing.

**Magnetometer calibration**

The measured magnetic field $B^{\text{meas}}(t)$ is determined from the output voltage of the LIA, denoted $V_{\text{LIA}}(t)$, by the relation $B^{\text{meas}}(t) = C_{\text{LIA}} V_{\text{LIA}}(t)$, where $C_{\text{LIA}}$ is a voltage-to-magnetic-field conversion factor given by

$$C_{\text{LIA}} = \frac{h}{\left.\frac{dV_{\text{LIA}}}{df}\right|_{V_{\text{LIA}}=0} g_e \, \mu_B \cos\left[\frac{\pi}{2} - \frac{\theta_{\text{tet}}}{2}\right]}.$$

Here $\left.\frac{dV_{\text{LIA}}}{df}\right|_{V_{\text{LIA}}=0}$ is the slope of the zero-crossing in V/Hz, $g_e$ is the electron g-factor, and $\mu_B$ is the Bohr magneton.

Calibration of the NV-diamond magnetometer was independently verified by applying a known test magnetic field $B_{\text{test}}(t) = B_{\text{test}} \text{square}[2\pi f_{\text{test}} t]$ with square wave amplitude $B_{\text{test}}$ and frequency $f_{\text{test}}$, and confirming the magnetometer records the correct value for $B^{\text{meas}}(t)$. The test magnetic field is produced by a multi-turn circular current loop (coil) with $N_{\text{turns}} = 7$ and radius $r_{\text{coil}} = 0.0235$ m, located a distance $z_{\text{coil}} = 0.103$ m from the diamond chip center. The coil is connected in series with an $R_{\text{series}} = 50\,\Omega$ resistor. The value of $B_{\text{test}}$ is calculated using the formula

$$B_{\text{test}} = \frac{\mu_0 N_{\text{turns}} I_{\text{coil}} r_{\text{coil}}^2}{2[z_{\text{coil}}^2 + r_{\text{coil}}^2]^{3/2}},$$

where $I_{\text{coil}}$ is the current in the coil generated by driving a voltage $V_{\text{coil}}(t)$ through the circuit. A 44 mV amplitude square wave yields $B_{\text{test}} = 1.8$ nT, with RMS voltage $B_{\text{test}}^{\text{rms}} = B_{\text{test}}$. When this value of $B_{\text{test}}$ is applied at frequency $f_{\text{test}} = 110$ Hz, the measured value of $B(t)$ is consistent with the value of $B_{\text{test}}$ to better than 5% as shown in Fig. S6a. A calibration without harmonics was also performed by applying a 62 mV amplitude sine wave yielding a consistent value of $B_{\text{test}}^{\text{rms}} = B_{\text{test}}/\sqrt{2} = 1.8$ nT.

**Magnetic field sensitivity**

A magnetometer's sensitivity is defined as $\eta = \delta B \sqrt{T}$, where $\delta B$ is the magnetic field signal that is as large as the noise, i.e., at SNR=1, after measurement time $T$[41]. The sensitivity of our NV-diamond magnetometer is evaluated using three methods. In method 1, a test magnetic field $B_{\text{test}}(t) = B_{\text{test}} \sin[2\pi f_{\text{test}} t]$ is applied for $N_{\text{trials}} = 150$, each of time $T_{\text{trial}} = 1$ s, and the measured magnetic field $B^{\text{meas}}(t)$ is recorded. For each trial $i$ the quantity

$$x_i = \frac{1}{T_{\text{trial}}} \int_0^{T_{\text{trial}}} B^{\text{meas}}(t) B_{\text{test}}(t)\, dt$$

is computed. The method 1 sensitivity $\eta_1$ is

$$\eta_1 = \frac{B_{\text{test}}^{\text{rms}} \sqrt{2}}{\mu} \sqrt{\frac{1}{N_{\text{trials}}} \sum_{i=0}^{N_{\text{trials}}} (x_i - \mu)^2} \times \sqrt{T_{\text{trial}}},$$

where $\mu \equiv \frac{1}{N_{\text{trials}}} \sum_{i=0}^{N_{\text{trials}}} x_i$, the factor of $\sqrt{2}$ accounts for in-quadrature noise, $B_{\text{test}}^{\text{rms}} = B_{\text{test}}/\sqrt{2}$, and typically $f_{\text{test}} = 250$ Hz. In method 2, $B_{\text{test}}(t)$ is applied for $N_{\text{trials}} = 150$, each of time $T_{\text{trial}} = 1$ s, and $B^{\text{meas}}(t)$ is recorded. The Fourier transform of $B^{\text{meas}}(t)$ is defined to be $\tilde{B}^{\text{meas}}(\omega) \equiv FFT[B^{\text{meas}}(t)]$. The method 2 sensitivity $\eta_2$ is

$$\eta_2 = B_{\text{test}}^{\text{rms}} \left\langle \frac{\frac{1}{f_{\text{stop}} - f_{\text{start}}} \int_{f_{\text{start}}}^{f_{\text{stop}}} |\tilde{B}^{\text{meas}}(2\pi f)|\, df}{\frac{1}{\Delta f} \int_{f_{\text{test}} - \Delta f/2}^{f_{\text{test}} + \Delta f/2} |\tilde{B}^{\text{meas}}(2\pi f)|\, df} \right\rangle_{N_{\text{trials}}} \times \sqrt{T_{\text{trial}}},$$

where $\Delta f = 1/T_{\text{trial}}$, the expected value is taken over $N_{\text{trials}}$, and typically $f_{\text{start}} = 300$ Hz, $f_{\text{stop}} = 600$ Hz, and $f_{\text{test}} = 250$ Hz. In method 3, no test magnetic field is applied and $B^{\text{meas}}(t)$ is recorded for $N_{\text{trials}} = 150$, each of time $T_{\text{trial}} = 1$ s; an example trace is shown in Fig. S6b. The sensitivity is then calculated as

$$\eta_3 = \sqrt{\frac{1}{T_{\text{trial}}} \int_0^{T_{\text{trial}}} [B^{\text{meas}}(t)]^2\, dt} \times \frac{1}{\sqrt{2 f_{\text{ENBW}}}},$$

with $f_{\text{ENBW}} = 4.0$ kHz. In all evaluations of the instrument's magnetic field sensitivity, $\eta_1 \sim \eta_2 \sim \eta_3$ was found, although $\eta_1$ converges most slowly and is therefore of limited use. Over 150 trials, $\eta_3$ ranges from 15.0 to 15.8 pT/$\sqrt{\text{Hz}}$, while $\eta_2$ is $15 \pm 1$ pT/$\sqrt{\text{Hz}}$. The two values are consistent. We thus conclude the NV-diamond magnetometer sensitivity is $15 \pm 1$ pT/$\sqrt{\text{Hz}}$, also consistent with a noise floor measurement of $|\tilde{B}^{\text{meas}}(2\pi f)|$ for $T_{\text{trial}} = 1$ s averaged over $N_{\text{trials}} = 150$, as shown in Fig. S6c, d.

This realized magnetic field sensitivity agrees with the expected sensitivity for our NV CW-ESR technique limited by photon shot noise and added MW and amplifier noise, as estimated herein. In the limit of low contrast $\mathcal{C}$ of the ODMR feature, the photon-shot-noise-limited sensitivity for CW-ESR magnetometry using NV-centers is given by[42]

$$\eta_{\text{ESR}} = \frac{4}{3\sqrt{3}} \frac{h}{g_e \mu_B} \frac{\Delta f}{\mathcal{C} \sqrt{\mathcal{R}}},$$

where $\mathcal{R}$ is the photon detection rate (away from resonance), $\Delta f$ is the power-broadened full-width-half-maximum (FWHM) resonance linewidth, and the factor $\frac{4}{3\sqrt{3}}$ comes from a Lorentzian feature's steepest slope. The CW-ESR method employed with the present NV-diamond magnetometer detects along two NV axes as described above, doubling the contrast while reducing the magnetic field sensitivity by the angle factor $\cos[\pi/2 - \theta_{\text{tet}}/2] = 0.8165$. The present instrument also uses modulation to reject quadrature noise, enhancing the sensitivity by an additional $\sqrt{2}$. As such, the shot-noise-limited sensitivity of our magnetometer is given nominally by

$$\eta_{\text{ESR}}^{\text{shot}} = \frac{1}{\sqrt{2}} \times \frac{4}{3\sqrt{3}} \frac{h}{g_e \mu_B} \frac{\Delta f}{\mathcal{C}_2 \cos\left[\frac{\pi}{2} - \frac{\theta_{\text{tet}}}{2}\right] \sqrt{\mathcal{R}}},$$

where $\Delta f = 1.5 \pm 0.1$ MHz is the measured linewidth; and $\mathcal{C}_2 = 5.3 \pm 0.1\%$ is the contrast when sensing along two NV axes, which was measured in the absence of modulation while addressing all three hyperfine features. The detected photon rate $\mathcal{R}$ is defined in terms of the photoelectron current $q\mathcal{R} = V_{\text{sig}}/R_L$, where $q$ is the elementary charge and $V_{\text{sig}} = 400$ mV is the typical signal photodiode voltage after $R_L = 50\,\Omega$ termination. This idealized shot-noise-limited CW-ESR sensitivity is found to be 2.9 pT/$\sqrt{\text{Hz}}$.

In practice, several factors diminish the sensitivity: first, the reference photodiode adds in quadrature an equivalent amount of shot noise, increasing the sensitivity by a factor $\mathcal{P}_{\text{ref}} = \sqrt{2}$; second, the slope is reduced with respect to the steepest slope of a Lorentizan due to the other nearby power-broadened hyperfine features, resulting in a sensitivity cost of $\mathcal{P}_{\text{slope}} = 1.19$. Taking these factors into account yields a shot–noise-limited CW-ESR sensitivity of 4.9 pT/$\sqrt{\text{Hz}}$.

Furthermore, in our square-wave modulated CW-ESR implementation, the contrast is reduced by an empirical factor $\mathcal{P}_{\text{mod}} \approx 1.6$, as shown in Fig. S6f, due to the finite cycling time of the NV center quantum states[43,44] and the loss of signal in higher harmonics resulting from demodulation with a sinusoidal lock-in frequency waveform[38]. The LNA-545 amplifier's noise figure of 1.8 increases the noise level by $\mathcal{P}_{\text{ampl}} \approx 1.23$. Application of MWs further increases the measured noise level by $\mathcal{P}_{\text{MW}} \approx 1.76$, as shown in Fig. S6e. These factors raise the expected magnetic field sensitivity to $\eta \approx \mathcal{P}_{\text{MW}} \mathcal{P}_{\text{ampl}} \mathcal{P}_{\text{mod}} \mathcal{P}_{\text{slope}} \mathcal{P}_{\text{ref}} \eta_{\text{shot}} \approx 17$ pT/$\sqrt{\text{Hz}}$, which agrees to within 13% of the measured $15 \pm 1$ pT/$\sqrt{\text{Hz}}$ for the data shown in Fig. S6e.

To confirm magnetometer sensitivity near the photon shot noise limit in the absence of applied MWs, we measured the RMS noise in $V_{\text{LIA}}$ for a range of power incident on the photodiode, at both 18 kHz and 90 kHz modulation frequencies, as shown in Fig. S6e. Data are fit to the function $y = (a + bx)^{1/c}$. For 90 kHz modulation, fit parameters are $a = (5.4 \pm 1.8) \times 10^{-6}, b = (3.0 \pm 0.6) \times 10^{-7}$, and $c = (1.97 \pm 0.04) \times 10^{-6}$, while for 18 kHz modulation the fit parameters are $a = (1.4 \pm 1.1) \times 10^{-5}, b = (4.4 \pm 3.0) \times 10^{-7}$, and $c = (1.90 \pm 0.14) \times 10^{-6}$. In both cases we observe $c \approx 2$, as expected for a shot-noise-limited measurement. The measured noise agrees with expected photoelectron shot noise plus LNA-545 amplifier noise for equivalent noise bandwidth $f_{\text{ENBW}} = 4.0$ kHz.

The fundamental sensitivity limit for spin-based magnetometers is given by the noise intrinsic to quantum projection. For a sample of $N$ electronic spins with characteristic dephasing time $T_2^*$, the spin-projection-noise-limited sensitivity is[45]

$$\eta_q = \frac{\hbar}{g_e \mu_B} \frac{1}{\sqrt{N T_2^*}}.$$

The sample used in this work has a total NV density $\sim 3 \times 10^{17}$ cm$^{-3}$ and no preferential orientation[46]. The density of NVs used to sense AP magnetic fields is reduced by a factor of two, as the AP magnetic field projects along only two NV axes. The illumination volume is $\sim 13\,\mu\text{m} \times 200\,\mu\text{m} \times 2\,\text{mm} \approx 5 \times 10^{-6}\,\text{cm}^3$, so the number of probed NV spins is $N \sim 8 \times 10^{11}$ with $T_2^* \approx 450$ ns. Using these values along with the electron gyromagnetic ratio $\gamma = g_e \mu_B / \hbar = 1.761 \times 10^{11}$ s$^{-1}$T$^{-1}$ gives a spin projection noise estimate for our sample of $\sim 10$ fT/$\sqrt{\text{Hz}}$. At $\sim 3000$ times better than the present nearly photon-shot-noise-limited sensitivity, there is much promise for significant gains in magnetometer sensitivity through use of pulsed magnetometry, optimized NV-diamond samples, and quantum-assisted techniques, as discussed below.

**Temporal resolution**
Temporal resolution of the NV-diamond magnetometer was tested by applying a test magnetic field $B_{\text{test}}(t) = B_{\text{test}}$ square$[2\pi f_{\text{test}} t]$ with $B_{\text{test}} \approx 57$ nT and $f_{\text{test}} = 1$ kHz, and measuring the 10% – 90% rise time of $B^{\text{meas}}(t)$, denoted by $\tau_{10/90}$. Using $f_{\text{mod}} = 60$ kHz, $\tau_{\text{LIA}} = 10\,\mu$s, and 6 dB/octave roll-off (yielding a measured $f_{\text{ENBW}} = 33$ kHz), $\tau_{10/90} = 32\,\mu$s is observed as shown in Fig. S7, which displays both real-time and averaged $B^{\text{meas}}(t)$ traces that are FFT low-pass filtered at 45 kHz. All AP data presented in this paper was acquired using $f_{\text{mod}} = 18$ kHz, $\tau_{\text{LIA}} = 30\,\mu$s, and 24 dB/octave roll-off, which gives $\tau_{10/90} \sim 400\,\mu$s. Note that higher values of $f_{\text{mod}}$ reduce NV spin-state contrast, an effect previously observed in refs. 41, 43, 44, and shown here in Fig. S6f. When operating with a temporal resolution higher than 40 µs, the magnetic field sensitivity of the present instrument is reduced by a factor of $\sim 1.6$ with respect to standard running conditions. With pulsed Ramsey-type schemes[22], to be employed in a next-generation NV-diamond magnetic imaging system, time resolution approaching the $\sim 200$ ns NV singlet state lifetime should be possible[47]. For example, in recent work to be published, we have shown that pulsed Ramsey schemes allow NV-diamond magnetic field measurements on $\sim 1\,\mu$s timescales.

**Expected magnetic field sensitivity in next-generation instrument**
A next-generation instrument will likely employ pulsed magnetic field sensing schemes, such as Ramsey-type sequences, which do not suffer from laser and MW power broadening of the ODMR features and thus allow for higher contrast than CW-ESR does[22]. A Ramsey scheme with free precession time $\tau$, and optical and MW initialization and readout times $t_I$ and $t_R$, has a shot noise sensitivity limit of[36,48,24]

$$\eta_{\text{Ramsey}} = \frac{\hbar}{g_e \mu_B} \frac{\sqrt{t_I + \tau + t_R}}{\tau} \frac{1}{\mathcal{C}\sqrt{\beta}},$$

where $\beta \approx \mathcal{R} t_R$ is the number of photons collected per measurement. Note that the contrast $\mathcal{C}$ is also dependent on $\tau$ due to spin dephasing. For $t_I, t_R \lesssim T_2^*$ the sensitivity is optimized for $\tau \sim T_2^*$. The optimal shot-noise-limited Ramsey sensitivity is then

$$\eta_{\text{Ramsey}} \sim \frac{\hbar}{g_e \mu_B} \frac{\sqrt{t_I + T_2^* + t_R}}{T_2^*} \frac{1}{\mathcal{C}\sqrt{\mathcal{R} t_R}}.$$

The principal improvements from employing a Ramsey scheme are in the contrast $\mathcal{C}$ and the lack of power broadening of ODMR features. Pulsed readout of the NV-diamond sample used in the present work realizes $\mathcal{C} = 9.5 \pm 0.5\%$ along a single NV axis. The CW-ESR power-broadened linewidth $\Delta f \approx 1.5$ MHz is also replaced by the natural linewidth $\frac{\Gamma}{2\pi} = \frac{1}{\pi T_2^*} \approx 700$ kHz. Note that there is a sensitivity cost in Ramsey schemes due to finite readout time, which can be estimated for typical values of $t_I = 1\,\mu$s and $t_R = 400$ ns, and using $\tau = T_2^* = 450$ ns for the current diamond sample. Assuming $\mathcal{R}$ remains the same as in the CW-ESR implementation because of finite available laser power, a net sensitivity improvement of about a factor of 5 is expected for pulsed magnetometry with respect to optimized CW-ESR. Moreover, because Ramsey sequences allow use of higher laser intensity than in CW-ESR, the sensitivity per illumination volume is further improved. With an expected $\sim 50$ times intensity increase, and thus a $\sim 50$ times smaller illumination volume for the same excitation power, an additional $\sim 50$ times enhancement in per-volume sensitivity is expected. Improved diamond samples with longer spin-dephasing times $T_2^*$ and higher nitrogen-to-NV conversion efficiencies will allow additional sensitivity gains. For the same $t_I$ and $t_R$, exchanging the present diamond for an equally bright diamond with $T_2^* = 32\,\mu$s would further improve the sensitivity approximately 17-fold. Moreover, quantum-beat magnetometry schemes[23,49] provide common-mode rejection of noise due to strain and temperature inhomogeneities, promising further sensitivity enhancement. Overall, through use of pulsed magnetometry and quantum-beat techniques with next-generation NV-diamond samples, a per-volume magnetic field sensitivity gain of $\sim 10^4$ should be possible.

**Simple magnetic model of action potential**
The magnetic field produced by an axon AP, denoted $B_{\text{axon}}(z, \rho, t)$, can be derived from the intracellular AP voltage $\Phi(z, \rho, t)$, where $z$ and $\rho$ denote the axial and radial coordinates respectively, using a simple model that agrees with more complex cable theory[50]. The axon is modelled as a conducting wire; hence the magnetic field is $B_{\text{wire}} = (\mu_0 I)/(2\pi \rho)$, with axial current $I$ due to the propagating AP. The wire's current density is $J = -\sigma \nabla \Phi(z, \rho, t)$, where $\sigma$ is the electrical conductivity. For a uniform cylindrical wire of radius $r_a$, the axial current may be expressed as $I = \pi r_a^2 J_z = -\pi r_a^2 \sigma \frac{\partial \Phi(z,\rho,t)}{\partial z}$. For constant values of conduction velocity $v_c$, the equality $\frac{\partial \Phi(z,\rho,t)}{\partial t} = -v_c \frac{\partial \Phi(z,\rho,t)}{\partial z}$ holds, where $v_c$ is defined to be positive[50]. Substitution then yields $I = \frac{\pi r_a^2 \sigma}{v_c} \frac{\partial \Phi(z,\rho,t)}{\partial t}$. Since $\Phi(z, \rho, t)$ is measured at a fixed point $z = z_0$, the partial time derivative can be replaced by a full derivative. At distances close to the axon surface where $\rho \sim r_a$, return currents outside the axon are minimal[50], and fringing effects from the finite axon length can be ignored, yielding

$$B_{\text{axon}}(z, \rho, t) = \frac{\mu_0 r_a^2 \sigma}{2 v_c \rho} \frac{d\Phi(z, \rho, t)}{dt}.$$

Defining $s \equiv \frac{\mu_0 r_a^2 \sigma}{2 v_c \rho}$ gives $B_{\text{axon}}(z, \rho, t) = s \frac{d\Phi(z,\rho,t)}{dt}$, where $s$ depends only on geometric and electrophysiological quantities. For the data shown in Fig. 2a-c, good proportionality is found between $B^{\text{meas}}(t)$ and $\frac{d\Phi^{\text{meas}}}{dt}$ with $s^{\text{meas}} = 7.6 \pm 1$ pT/(V/s).

Accurate calculation of $s$ from first principles is nontrivial[51], since $r_a$, $\rho$, $\sigma$, and $v_c$ have substantial uncertainties. $r_a = 200 \pm 75\,\mu$m is determined from stained transverse sections of *M. infundibulum* (see Fig. 4), with large variations observed in axon size (up to 50%) among otherwise similarly sized specimens, as also noted in[19]. For the excised axon studies, only the ventral nerve cord containing the giant axon is isolated, and there is also residual connective tissue around the axon: hence an estimate of $\rho = r_a + 100\ \mu m \pm 100\ \mu m$ is used. We take $\sigma = 1.47 \pm 0.5$ S/m[52], given the significant variation (50%) in axoplasm conductivity reported for *L. pealeii*[51]; and $v_c = 9 \pm 4$ m/s based on two-point electrophysiology measurements of $v_c$ for similar representative-sized worms under posterior stimulation. With these values for $r_a$, $\rho$, $\sigma$, and $v_c$, we extract $s^{\mathrm{calc}} = 13.7 \pm 10$ pT/(V/s), which is in agreement with the experimentally derived value.

To evaluate the feasibility of NV-diamond magnetic sensing of small mammalian neurons, a crude estimate of the AP magnetic signal size was made for Purkinje neurons using our simple model. We used $\sigma = .66\ \Omega^{-1}\ m^{-1}$, an average of the values $.44\ \Omega^{-1}\ m^{-1}$ from ref. 53, $.87\ \Omega^{-1}\ m^{-1}$ from ref. 54, and $.67\ \Omega^{-1}\ m^{-1}$ from ref. 55. We used $\frac{d\Phi}{dt} = 339$ V/s, an average of 300 V/s from ref. 56, 367 V/s from ref. 57, and 350 V/s from ref. 58. We used $v_c = .25$ m/s, an average of .24 m/s from ref. 56 and .25 m/s from ref 59. For $r_a = 1\ \mu m, 2\ \mu m$, and $3\ \mu m$, we calculate a peak magnetic field of $B_{\mathrm{axon}}^{\max} = .6$ nT, 1.1 nT and 1.7 nT respectively at the axon surface. This calculation is intended for rough estimation purposes only. We acknowledge that conduction velocity is expected to be correlated with diameter and this is not accounted for in this calculation.

**Action potential signal-to-noise ratio**
The SNR of an AP magnetic field data set is calculated using (i) the peak-to-peak detected AP signal from an averaged set of $N_{\mathrm{avg}}$ measurements and (ii) the standard deviation of the time trace in a section of the same data set in which no AP is present. The single-shot SNR is calculated by dividing the SNR of the averaged data by $\sqrt{N_{\mathrm{avg}}}$. For excised axon studies, $t = 0$ corresponds to the beginning of the stimulation pulse. For averaging data in intact organism studies, traces are aligned in time using a digital trigger set on either the maximum or minimum of the extracellular AP voltage signal $\Phi_{\mathrm{ex}}(t)$; this alignment compensates for specimen contractions and thus prevents smearing out of the averaged signal. To maximally improve the SNR of a known expected signal in the presence of stochastic noise, it can be shown that a matched filter is the optimal linear filter. For a detected signal $x(t)$ containing an expected signal and additive noise, the matched filtered signal $y(t)$ is given by the convolution

$$y(t) = \int_0^t h(t - t')x(t')dt',$$

where $h(t)$ is the time-reversed trace of the expected signal. The matched filter for the data shown in Fig. 2c was constructed from a larger set of 600 detected AP time traces all from the same organism. The average of the 600 traces was high-pass FFT filtered at 80 Hz to prevent non-DC values due to drift from being interpreted as signal. The trace was then zeroed for all times except a 1.4 ms window that includes the full detected AP signal, time-reversed, and then taken as the expected signal $h(t)$ for the matched filter. This filter was applied to the four consecutive sets of 150 averages contained in the larger data set. The SNR of each of these filtered traces was improved to be between 14.5 and 16, indicating that the SNR of a single AP event after filtering is $1.2 \pm 0.1$.

**Systematic checks**
Multiple tests were performed, as summarized in Table S2, to verify that the observed $B^{\mathrm{meas}}(t)$ arises solely from an axon AP (i.e., intracellular axial current): (i) observation of a non-zero $B^{\mathrm{meas}}(t)$ signal required successful AP stimulation and propagation as determined by electrophysiology measurements of the extracellular action potential $\Phi_{\mathrm{ex}}(z, \rho, t)$; (ii) crosstalk ('pickup artifacts') during data acquistion between the recorded $\Phi_{\mathrm{ex}}(z, \rho, t)$ and $V_{\mathrm{LIA}}(t)$ was ruled out by varying the recording electrode placement and observing no change in $B^{\mathrm{meas}}(t)$; (iii) the origin of the NV-observed $B^{\mathrm{meas}}(t)$ signal was demonstrated to be magnetic by switching to an LIA voltage zero-crossing with slope $\frac{dV_{\mathrm{LIA}}}{df}$ of opposite sign, and observing inversion of $B^{\mathrm{meas}}(t)$; (iv) similarly, inverting the phase of the LIA reference signal $\phi_{\mathrm{LIA}}$ by 180° produced the same result, also confirming the magnetic origin of the signal sensed by the NV ensemble; and (v) time-varying magnetic fields from motional artifacts, e.g. specimen-induced instrument motion in the presence of a gradient in the bias field $B_0$, were ruled out by reversing the orientation of the permanent magnet and observing inversion of $B^{\mathrm{meas}}(t)$.

**Directional detection of action potentials**
For three specimens, (Worms C, D, and E in Fig. S8a), magnetic AP signals were recorded for both posterior stimulation and anterior stimulation, each for $N_{\mathrm{avg}} = 1650$ trials. The measured magnetic signal $B^{\mathrm{meas}}(t)$ was inverted for anterior stimulation compared to posterior stimulation, as expected. In addition, larger peak-to-peak values of $B^{\mathrm{meas}}(t)$ were observed for posterior stimulation than for anterior stimulation for all three worms tested, by $47\% \pm 20\%$. This result did not depend on which stimulation (posterior or anterior) was tested first; and was robust under multiple switches of stimulation (e.g., posterior, anterior, then posterior again).

To confirm the directional dependence of the AP conduction velocity, electrophysiology recordings were perfomed simultaneously at two points separated by 6 to 10 mm along the whole worm giant axon using two sets of bipolar measurement electrodes. The electrodes were connected to a differential amplifier (A-M Systems Model 1800 Headstage), which was further amplified (A-M systems Model 1800) and digitized through an oscilloscope (Tektronix TDS2004B). The delay between the initialization of stimulation and the peak AP signal on each pair of electrodes was measured, and the conduction velocity was determined from the timing difference between the detected signals and the spatial separation between the electrode pairs.

**Extended duration action potential sensing**
For the long-term sensing data shown in Fig. S8b, the specimen (Worm F) was prepared and clamped to the apparatus as described above for intact organism studies. The worm was magnetically monitored for $> 24$ hours in the presence of applied MWs and laser illumination of the diamond. Following this duration, the magnetic AP signal $B^{\mathrm{meas}}(t)$ was measured to have an amplitude consistent with AP signals of specimens studied over shorter durations (Worms C, D, and E). Physical stimulus applied to the worm further confirmed its responsiveness and health.

| Technology | Single-neuron scale | Whole-organism scale | Labelling required | Invasive | Phototoxic | Long-term measurement stability | Imaging | Spatial resolution | Super-resolution | Field of view | Sensing depth | Temporal resolution | Single event detection | Measures conduction velocity | Measures AP propagation direction |
|---|---|---|---|---|---|---|---|---|---|---|---|---|---|---|---|
| Traditional intracellular electrophysiology | YES | NO | NO | YES | NO | NO[5] | NO | n/a | n/a | n/a | > 1 mm[60] | ~50 μs[61] | YES | YES[5] | YES[5] |
| All-optical electrophysiology | YES | YES | YES | YES[62] | YES[62] | NO[62] | YES | Diffraction-limited | NO | > 4 mm[62] | ~1 mm[1] | ~100 μs[63] | YES | YES[62] | YES[62] |
| Calcium imaging | YES | YES | YES | YES[64] | YES[65] | NO[65] | YES | Diffraction-limited | NO | 0.5 mm[66] | ~1 mm[1] | 100 ms[64] | YES | NO[64] | NO[64] |
| Microelectrode array | YES | NO | NO | NO | NO | YES[67] | YES | ~10 μm | n/a | 1 cm[6] | ~1 cm[68] | ~50 μs[69] | YES | YES[69] | YES[69] |
| SQUID (Wikswo technique)[51] | YES | NO | NO | NO | NO | Not demonstrated | NO | n/a | n/a | n/a | ~1 cm[70] | ~50 μs[51] | YES | YES[51] | Not demonstrated |
| fMRI | NO | YES | NO | NO | NO | YES | YES | ~1 mm | NO[71] | > 10 cm[72] | > 10 cm[73] | ~1 s[74] | NO | n/a | n/a |
| MEG (uses SQUID)[75] | NO | YES | NO | NO | NO | YES | YES | ~1 mm | NO | > 10 cm[75] | > 10 cm[75] | ~1 ms[75] | NO | n/a | n/a |
| **Demonstrated** NV-diamond magnetic imaging | YES | YES | NO | NO | NO | YES | YES | ~10 nm[3] | YES | ~1 mm[4] | ~1 mm | ~30 μs | NO | YES | YES |
| **Projected** NV-diamond magnetic imaging | YES | YES | NO | NO | NO | YES | YES | ~10 nm | YES | ~1 cm | ~1 cm | ~1 μs | YES | YES | YES |

**Table S1 | Competing technology comparision**

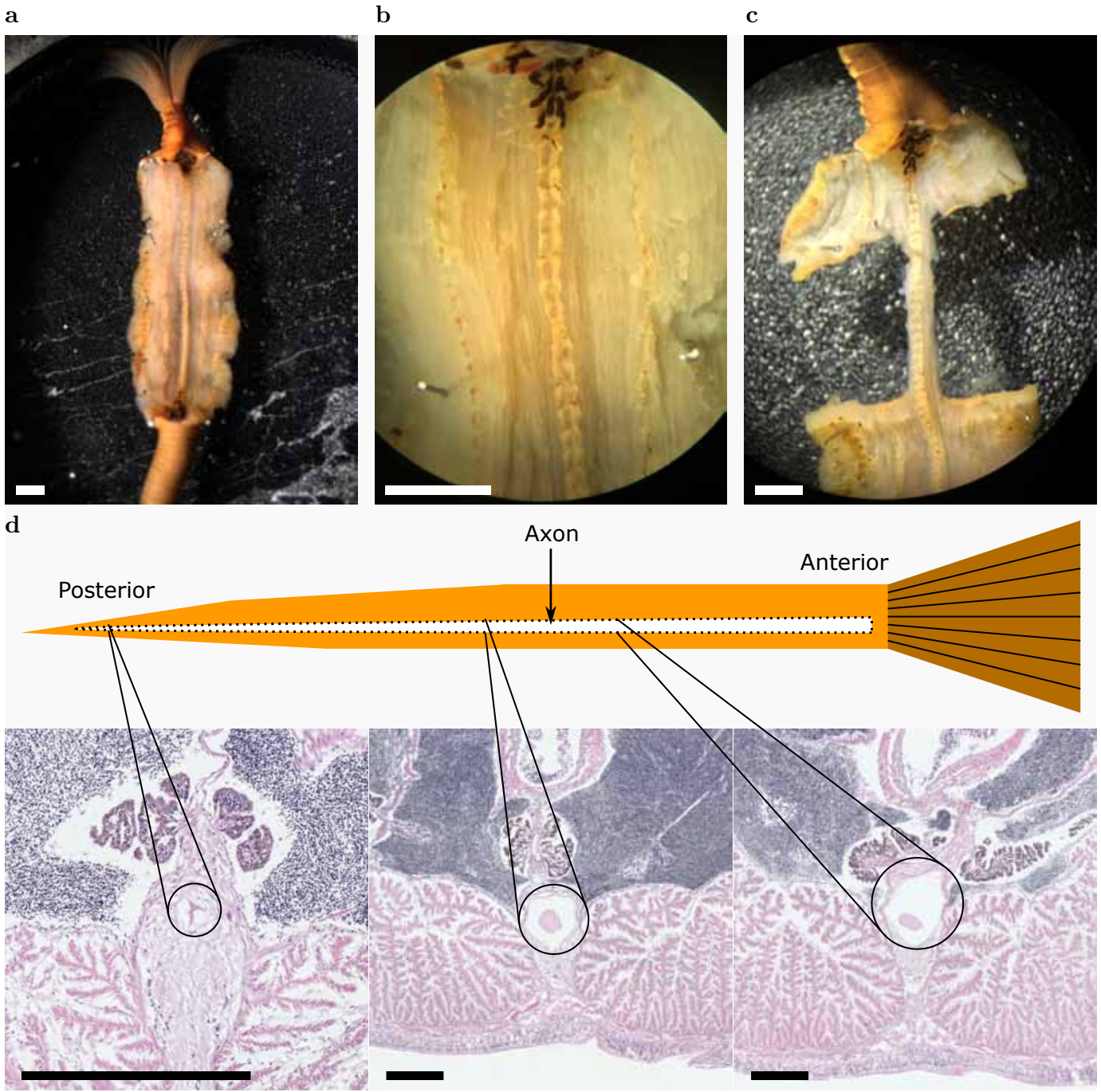

**Figure S1 | Dissected specimen of *M. infundibulum*.** **a,** Photo of worm with nerve cord containing giant axon exposed as discussed in Methods. **b,** Close-up view of same specimen. The nerve cord is ~ 400 μm across near the anterior (top) end. **c,** Same specimen with muscle tissue removed. At this stage the dissection protocol is complete. All white scale bars correspond to 2 mm. The levels of each photo were slightly and uniformly adjusted for improved contrast. **d,** Cartoon drawing of worm and transverse sections. Middle and right sections are reproduced from Fig. 4 a, b; leftmost section is from near the tip of the posterior end of the specimen, further demonstrating the significant tapering of the giant axon. All black scale bars correspond to 400 μm.

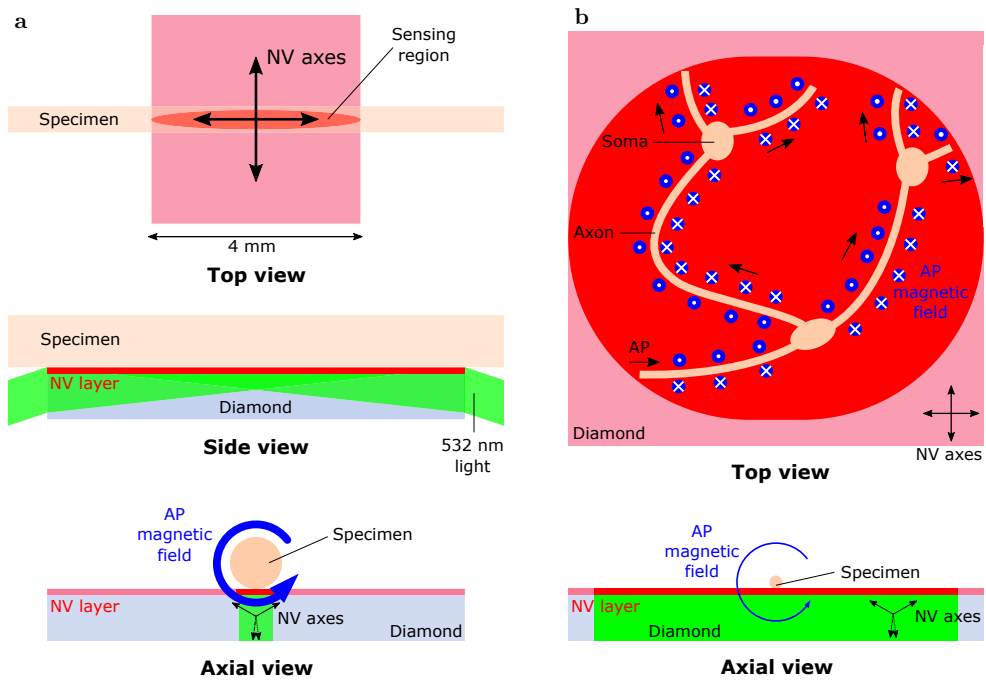

**Figure S2 | Specimen orientation with respect to NV axes. a,** Present specimen orientation as discussed in Methods and main text. Diagram is reproduced from Fig. 1 for comparison. **b**, Proposed method for magnetic imaging of AP dynamics from networks of smaller neurons with arbitrary orientation. Here the sensor detects the magnetic field component normal to the diamond surface, which has opposite sign on different sides of the specimen.

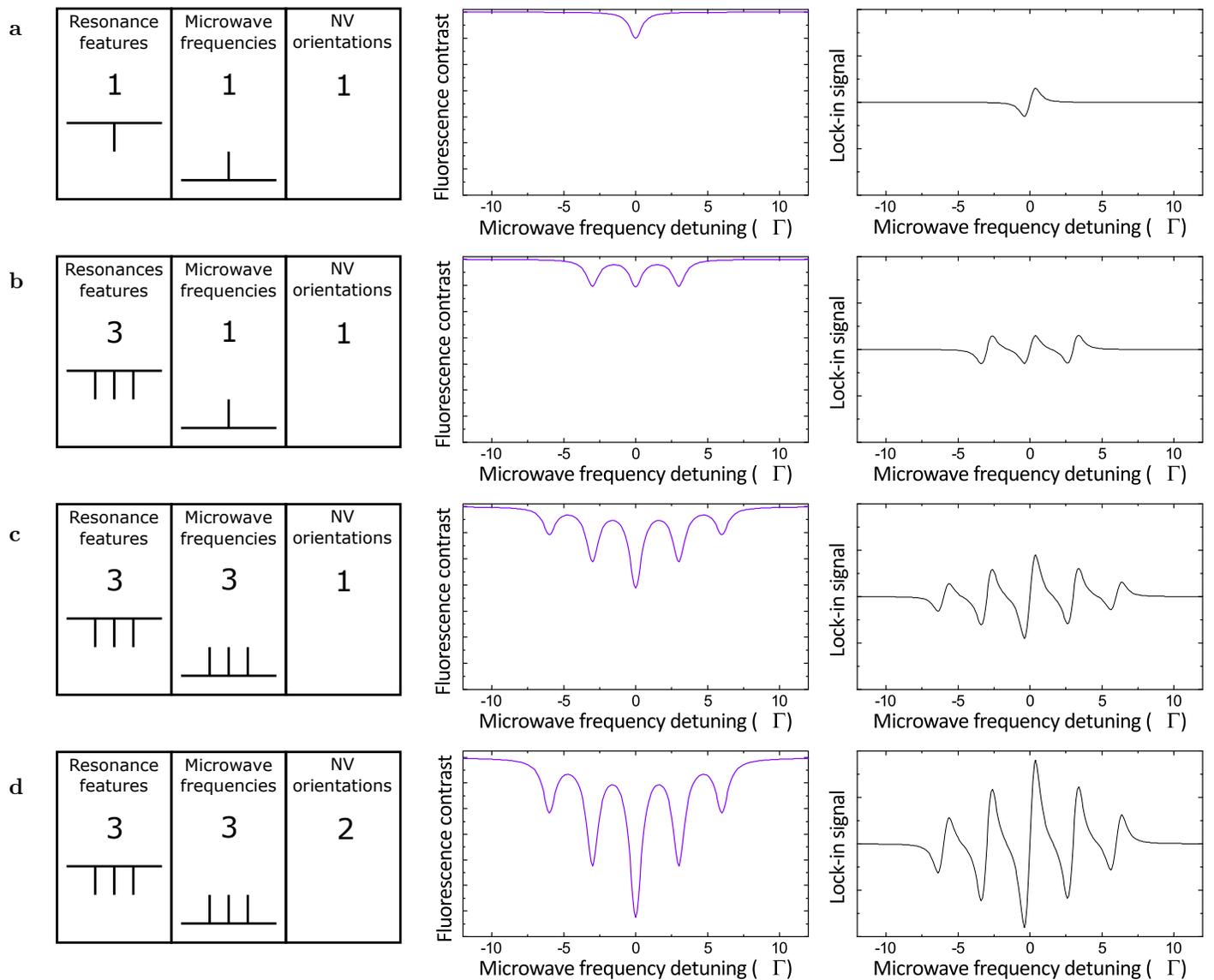

**Figure S3 | Illustration of magnetometry technique.** Left column shows schematic diagrams illustrating number of ODMR features, number of MW frequencies applied, and number of NV axes used for sensing; middle column shows calculated ODMR fluorescence profiles in units of the natural linewidth Γ, in the absence of power broadening; and right column shows associated dispersion-type lock-in amplifier (LIA) signals. Fluorescence and LIA signals are given in arbitrary units. See Methods for discussion of the lock-in scheme. **a**, Diagram, fluorescence signal, and LIA signal for a single ODMR feature addressed by a single (modulated) MW frequency, sensed along a single NV axis. **b**, Diagram, fluorescence signal, and LIA signal for three ODMR features addressed by a single (modulated) MW frequency, sensed along a single NV axis. **c**, Diagram, fluorescence signal, and LIA signal for three ODMR features addressed by three (modulated) MW frequencies with equivalent spacing, sensed along a single NV axis. The central feature corresponds to all three applied frequencies resonantly addressing ODMR features, as described in Methods. **d**, Diagram, fluorescence signal, and LIA signal for same scenario as in (c) but with $B_0$ oriented to have equal projection along two NV axes, overlapping their ODMR features, as discussed in Methods.

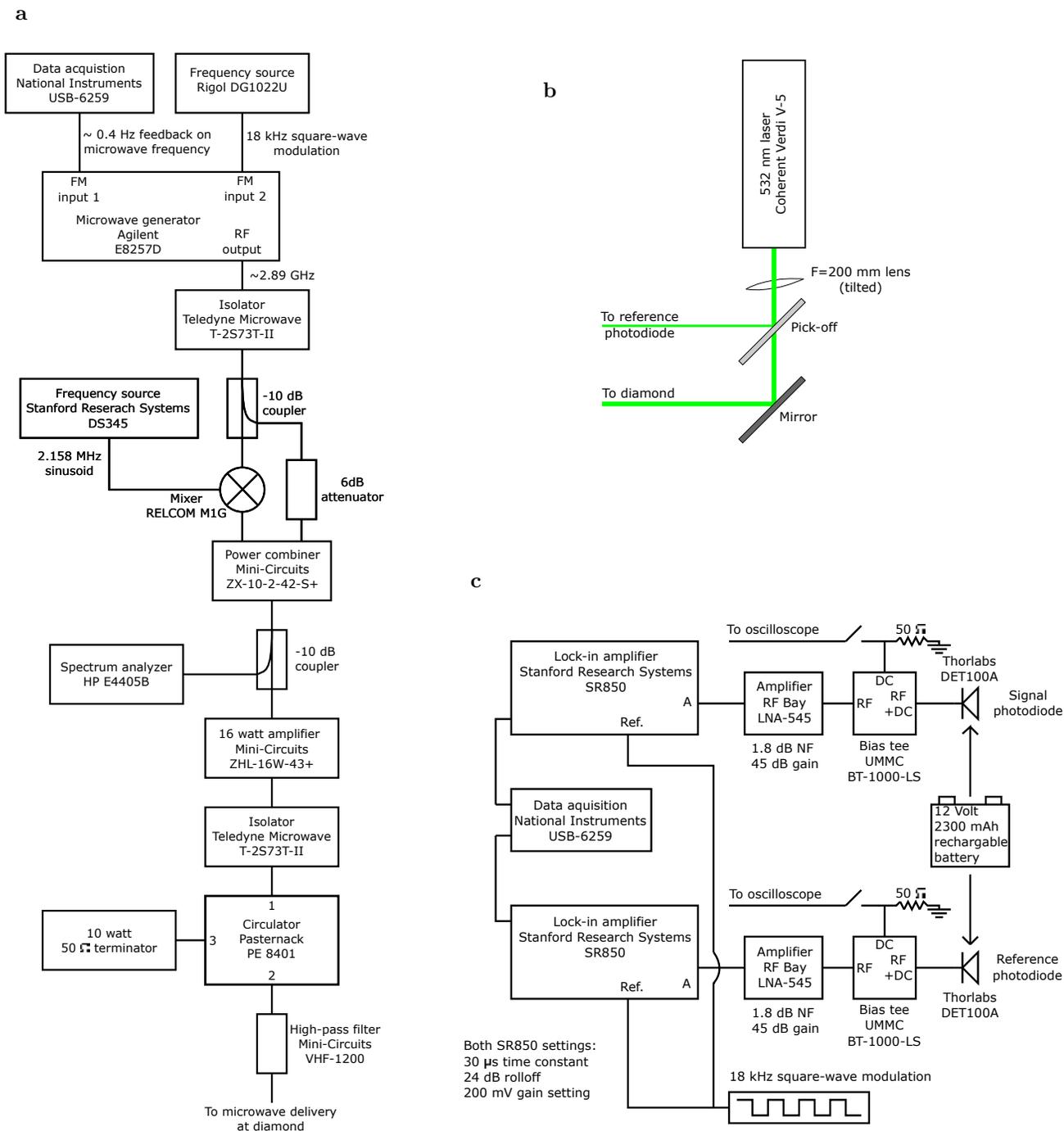

**Figure S4 | Microwave, laser, and light collection setup.** **a,** MW generation, modulation, and delivery setup as described in Methods. **b,** Laser setup as described in Methods. **c**, Signal photodiode, reference photodiode, and downstream electronics and LIAs as described in Methods.

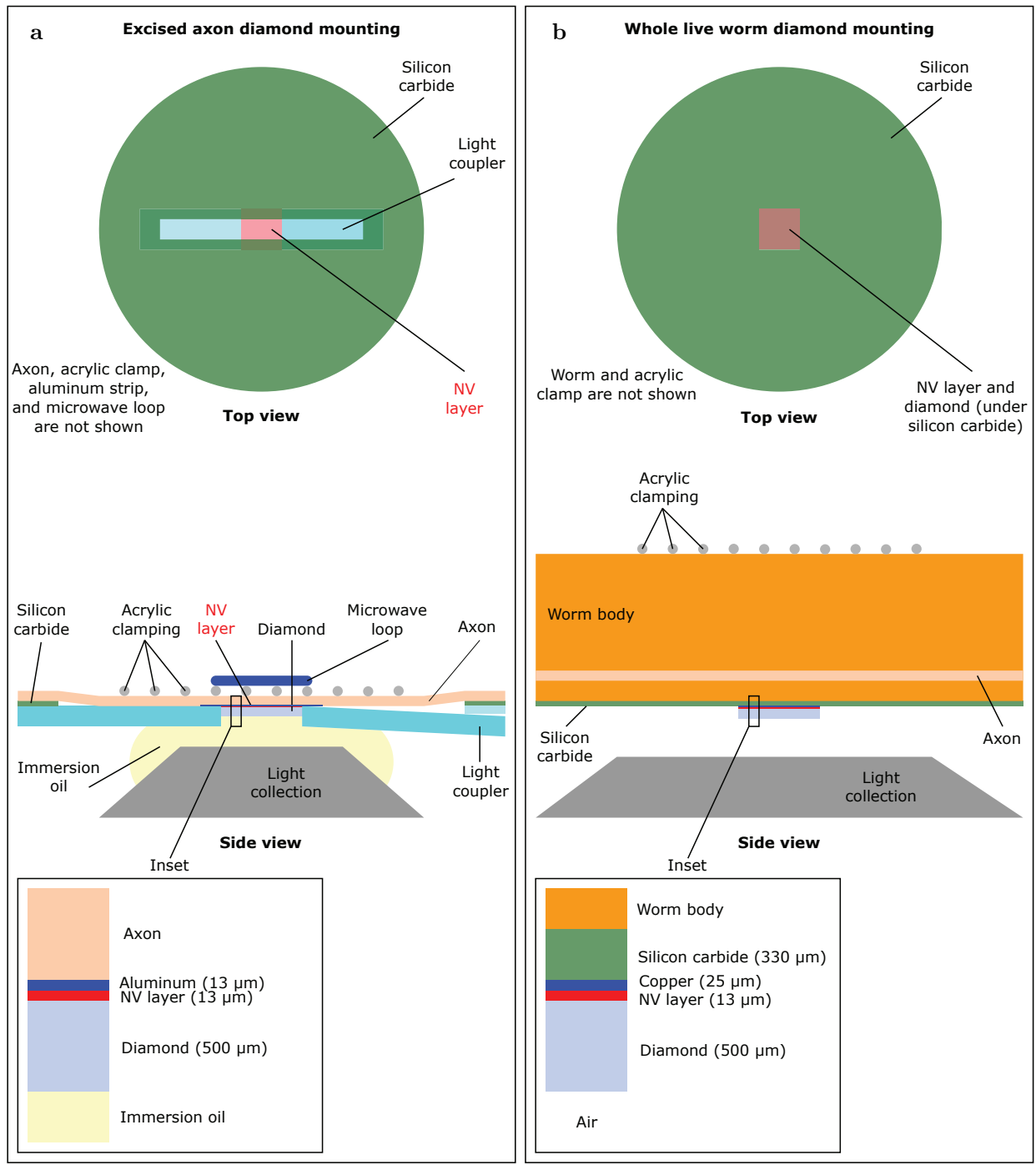

**Figure S5 | Diamond mounting. a,** Diamond mounting for excised axon experiments as described in Methods and main text. **b,** Diamond mounting for whole live worm experiments as described in Methods and main text. MWs are applied through the 25 μm thick copper layer.

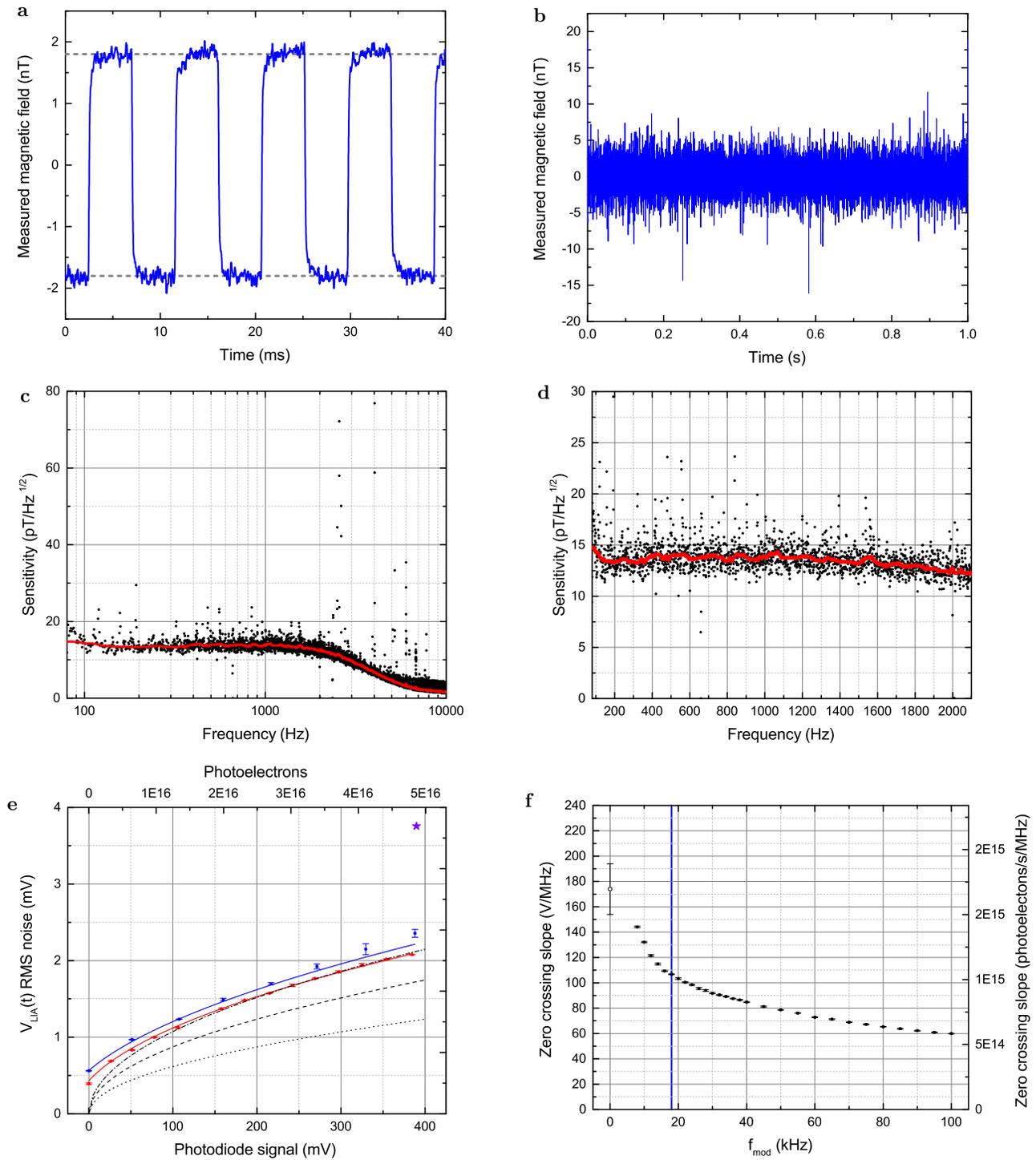

**Figure S6 | Magnetometer calibration and sensitivity.** **a**, Calibration verification as described in detail in Methods. A 110 Hz square wave with 1.8 nT amplitude (calculated from coil geometry, coil distance to diamond sensor, and current through coil only) is averaged for $N_{avg} = 1000$ trials. The measured magnetic field, calibrated only from the value of $C_{LIA}$ and the lock-in amplifier voltage time trace $V_{LIA}(t)$, is consistent with a 1.8 nT amplitude square wave to better than 5%. Gray dashed lines depict -1.8 nT and +1.8 nT levels. The slight rounding of the square wave's corners results from coil non-idealities rather than the magnetometer. **b**, Real-time trace of measured magnetic field $B^{meas}(t)$ with no external time-varying magnetic field applied. **c**, Fourier transform (black points) of $N_{avg} = 150$ traces of (b) is smoothed (red line) for clarity and is consistent with an overall sensitivity of 15 pT/$\sqrt{Hz}$. **d**, Reproduction of (c) with linear scale over approximate neuron signal bandwidth (80 Hz to 2 kHz). All data are taken for standard conditions ($f_{mod} = 18$ kHz, nominal $\tau_{LIA} = 30$ μs, with 24 dB/octave roll-off). **e,** Measured and calculated RMS noise on $V_{LIA}(t)$ versus PD signal voltage. Data shown in blue (red) were taken without applied MWs at $f_{mod} = 18$ (90) kHz. Blue and red curves are fits to the respective data sets, (discussed in Methods), demonstrating the square-root dependence of the measured noise. Purple star marks the measured noise during typical operating conditions in the presence of applied MWs. Black curves indicate calculated theoretical noise for shot noise from a single channel (dotted), shot noise including both the signal and reference channels (dashed), and expected noise level including shot noise from both channels and the LNA-545 amplifier noise figure of 1.8 (dot-dashed). **f,** Measured slope of the zero-crossing $\frac{dV_{LIA}}{df}\Big|_{V_{LIA}=0}$ with modulation frequency $f_{mod}$. Blue line denotes $f_{mod} = 18$ kHz. The open circle marks the slope in the absence of modulation, calculated from the measured DC photodiode signal and the LNA-545 amplifier and LIA gains.

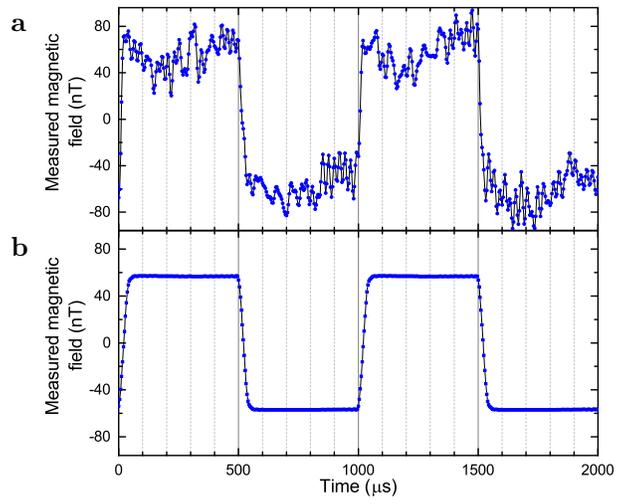

**Figure S7 | Temporal resolution. a,** Real-time trace of 1 kHz square wave with 57 nT amplitude as described in Methods. **b**, Same experimental setup as (a), but highly averaged ($N_{avg} \sim 10^6$). Data analysis indicates a 10% - 90% rise time of $\tau_{10/90} = 32$ µs. For this data only, $f_{mod} = 60$ kHz, $\tau_{LIA} = 10$ µs nominally, and a 6 dB/octave roll-off is used, yielding a measured $f_{ENBW} = 33$ kHz. Data are FFT low-pass filtered at 45 kHz.

| Reversal | Result | Systematic ruled out |
|---|---|---|
| Axon firing → Axon not firing | $B^{\text{meas}}(t) \to 0$ | Any non-AP signal, including stimulation artifacts |
| $\Phi_{\text{ex}}(z_1, \rho_1, t) \to \Phi_{\text{ex}}(z_2, \rho_2, t)$ | $B^{\text{meas}}(t) \to B^{\text{meas}}(t)$ | Cross-talk of $B^{\text{meas}}(t)$ with $\Phi_{\text{ex}}(z, \rho, t)$ |
| $\text{sgn}[\frac{dV_{\text{LIA}}}{df}] = 1 \to \text{sgn}[\frac{dV_{\text{LIA}}}{df}] = -1$ | $B^{\text{meas}}(t) \to -B^{\text{meas}}(t)$ | Any non-magnetic artifact including cross-talk of $B^{\text{meas}}(t)$ with $\Phi_{\text{ex}}(z, \rho, t)$ |
| $\phi_{\text{LIA}} \to \phi_{\text{LIA}} + 180°$ | $B^{\text{meas}}(t) \to -B^{\text{meas}}(t)$ | Cross-talk of $B^{\text{meas}}(t)$ with $\Phi_{\text{ex}}(z, \rho, t)$ |
| $B_0 \to -B_0$ | $B^{\text{meas}}(t) \to -B^{\text{meas}}(t)$ | Magnetic artifact from motional coupling to small $\frac{\partial B_0}{\partial x}$ across diamond |

**Table S2 | Systematic checks**

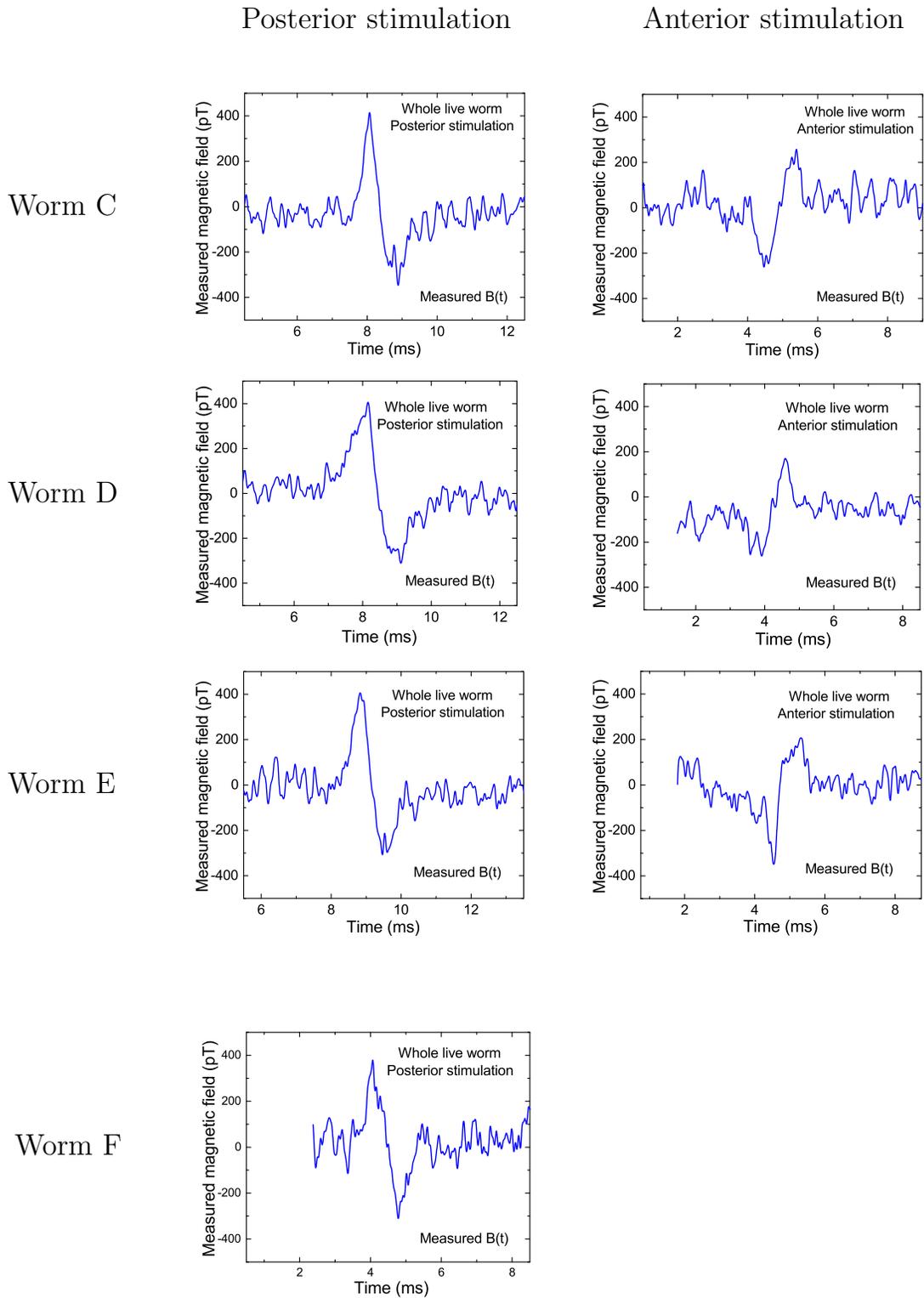

**Figure S8 | Directional sensing for three specimens and extended duration sensing. a,** Measured time trace of AP magnetic field $B^{\text{meas}}(t)$ for *M. infundibulum* giant axons with $N_{\text{avg}} = 1650$. Data are shown for Worms C, D, and E for both posterior stimulation (left column) and anterior stimulation (right column). For the sample consisting of Worms C, D, and E, the peak-to-peak value of $B^{\text{meas}}(t)$ is larger by 47% ± 20% for posterior stimulation than for anterior stimulation. **b,** Measured time trace of AP magnetic field $B^{\text{meas}}(t)$ for *M. infundibulum* giant axon (Worm F) with $N_{\text{avg}} = 1200$. Worm F was continuously magnetically monitored with full laser and MW power for 24 hours prior to stimulation and measurement of the AP magnetic field $B^{\text{meas}}(t)$. The peak-to-peak value of $B^{\text{meas}}(t)$ for worm F is not statistically different from the peak-to-peak value of $B^{\text{meas}}(t)$ for worms C, D, or E with posterior stimulation. These results indicate little if any negative effects from this sensing method over long time periods.